%% file: main.tex
\documentclass[letterpaper,twocolumn,10pt]{article}
\usepackage{./usenix-2020-09}

\usepackage{amsmath}

\usepackage{amssymb,amsfonts}
\usepackage{algorithmic}
\usepackage{graphicx}
\usepackage{textcomp}
\usepackage{standalone}
\usepackage{xcolor}
\def\BibTeX{{\rm B\kern-.05em{\sc i\kern-.025em b}\kern-.08em
    T\kern-.1667em\lower.7ex\hbox{E}\kern-.125emX}}
\usepackage[utf8]{inputenc}
\usepackage{booktabs} 
\usepackage{graphicx}
\usepackage{subfigure}
\usepackage{units}
\usepackage{blindtext}
\usepackage{multirow}
\usepackage{xcolor,xspace}
\usepackage{amsmath}
\usepackage{paralist}
\usepackage{mdframed}
\usepackage{pict2e}
\usepackage{mdframed}
\usepackage{listings}
\usepackage{color}
\usepackage{threeparttable}
\usepackage{array}
\usepackage{booktabs}
\usepackage{pifont}
\usepackage{url}
\usepackage{comment}
\usepackage{amsmath}
\usepackage{lipsum}
\usepackage{arydshln}
\usepackage{makecell}
\usepackage{flushend}

\usepackage{latexsym}
\usepackage{enumitem}
\usepackage{arabtex}
\usepackage{fancyhdr}
\usepackage[english]{babel}
\usepackage{tabularx}
\input{glyphtounicode}

\usepackage[ruled]{algorithm2e}
\usepackage{algorithmic}

\usepackage{url}

\usepackage[
    n,
    advantage,
    operators,
    sets,
    adversary,
    landau,
    probability,
    notions,
    logic,
    ff,
    mm,
    primitives,
    events,
    complexity,
    asymptotics,
    keys]{cryptocode}

\usepackage{tikz}                    
\usetikzlibrary{arrows, patterns, automata, positioning}

\newcommand{\ignore}[1]{}

\newcommand{\edit}[1]{\textcolor{blue}{#1}}

\newcommand{\prv}{{\ensuremath{\sf{\mathcal Prv}}}\xspace}
\newcommand{\vrf}{{\ensuremath{\sf{\mathcal Vrf}}}\xspace}
\newcommand{\RA}{{\ensuremath{\sf{\mathit{RA}}}}\xspace}
\newcommand{\PoX}{{\ensuremath{\sf{\mathit{PoX}}}}\xspace}

\newcommand{\chal}{\ensuremath{Chal}\xspace}

\newcommand{\CFA}{{\textit{CFA}}\xspace}

\newcommand{\acron}{\textit{{ACFA}}\xspace}
\newcommand{\vrased}{{VRASED}\xspace}
\newcommand{\garota}{{GAROTA}\xspace}
\newcommand{\tcb}{\textit{TCB}\xspace}
\newcommand{\s}{\textit{S}\xspace}
\newcommand{\ptr}{\mathit{CF_{size}} \xspace}

\newcommand{\trigger}{{\ensuremath{\sf{trigger}}}\xspace}

\newcommand{\inst}{{\textit{inst}}\xspace}

\newcommand{\attkey}{\ensuremath{\mathcal K}\xspace}
\newcommand{\auth}{\ensuremath{\mathcal Auth}\xspace}

\renewcommand\adv{\ensuremath{\sf{\mathcal Adv}}\xspace}

\newcommand{\cfattest}{\ensuremath{\mathtt{HMAC}}\xspace}

\newcommand{\vrfy}{\ensuremath{\mathtt{Verify}}\xspace}
\newcommand{\authen}{\ensuremath{\mathtt{Authenticate}}\xspace}

\newcommand{\func}{\ensuremath{\mathcal{F}}\xspace}

\newcommand{\er}{\textit{AER}\xspace}

\renewcommand{\log}{\ensuremath{\mathcal{CF}_{Log}}\xspace}

\newcommand{\reqchange}[2]{#2}
\newcommand{\blue}[1]{#1}

\usepackage[available,functional,reproduced]{usenixbadges}

\begin{document}
%
\title{ACFA: Secure Runtime Auditing \& Guaranteed Device Healing\\ via Active Control Flow Attestation}

\author{
{\rm Adam Caulfield}\\
Rochester Institute of Technology 
\and
{\rm Norrathep Rattanavipanon}\\
Prince of Songkla University, Phuket Campus
\and
{\rm Ivan De Oliveira Nunes}\\
Rochester Institute of Technology 
} 

\maketitle

\begin{abstract}
Embedded devices are increasingly used in a wide range of ``smart'' applications and spaces.
At the lower-end of the scale, 
they are implemented under strict cost and energy budgets, using microcontroller units (MCUs) that lack  security features akin to those available in general-purpose processors.
In this context, Remote Attestation (\RA) was proposed as an inexpensive security service that enables a verifier (\vrf) to remotely detect illegal modifications to the software binary installed on a prover MCU (\prv). Despite its effectiveness to validate \prv's binary integrity, attacks that hijack the software's control flow (potentially leading to privilege escalation or code reuse attacks) cannot be detected by classic \RA.

Control Flow Attestation (\CFA)
augments \RA with information about the exact order in which instructions in the binary are executed. As such, \CFA enables detection of the aforementioned control flow attacks.
However, we observe that current \CFA architectures cannot guarantee that \vrf ever receives control flow reports in case of attacks. In turn, while they support detection of exploits, they provide no means to pinpoint the exploit origin. Furthermore, existing CFA requires either (1) binary instrumentation, incurring significant runtime overhead and code size increase; or (2) relatively expensive hardware support, such as hash engines. In addition, current techniques are neither continuous (they are only meant to attest small and self-contained operations) nor active (once compromises are detected, they offer no secure means to remotely remediate the problem).

To jointly address these challenges, we propose \acron: a hybrid (hardware/software) architecture for \underline{\textit{A}}ctive \underline{\CFA}. \acron enables continuous monitoring of all control flow transfers in the MCU and does not require binary instrumentation. It also leverages the recently proposed concept of ``active roots-of-trust'' to enable secure auditing of vulnerability sources and guaranteed remediation, in case of compromise detection.
We provide an open-source reference implementation of \acron on top of a commodity low-end MCU (TI MSP430) and evaluate it to demonstrate its security and cost-effectiveness.
\end{abstract}

\maketitle

\section{Introduction}
\label{sec:intro}

Embedded devices are crucial components of modern systems.
A large portion of these devices are implemented using low-end and bare-metal microcontroller units (MCUs), specifically designed for energy, cost, and spatial efficiency. They are well-suited for and commonly used in safety-critical sensor-based applications such as medical devices, vehicular sensors/actuators, and sensor/alarm systems. Due to cost/energy budgets, MCUs often lack common hardware features used to secure higher-end systems (e.g., memory management units, strong privilege separation, and inter-process isolation). 
Unsurprisingly, the absence of these features makes them attractive targets to a wide range of software attacks~\cite{deogirikar2017security,stuxnet,giraldo2016integrity}.

In this context, Remote Attestation (\RA)~\cite{smart,vrased,tytan,trustlite,simple,hydra,rata,DAA,Sancus17,scraps,pistis,reserve,sacha}, as well Proofs of Execution (\PoX)~\cite{apex, asap}, have been proposed as means for a verifier (\vrf) to ascertain the software state of a remote prover MCU (\prv). 
While these techniques can prove software integrity and its execution on \prv , they cannot detect runtime attacks that tamper with the program's control flow without modifying its code.
For instance, an adversary (\adv) can leverage a buffer overflow vulnerability to hijack the program's control flow by overwriting the return address of the executing function. This vulnerability can in turn be used to jump to an arbitrary instruction within the binary, potentially skipping security checks or launching Return-Oriented Programming (ROP)~\cite{rop} attacks.

Control Flow Integrity (CFI) methods~\cite{FC08,hafix,DSL14} 
aim to address these vulnerabilities by proactively checking the program's control flow at runtime, locally at \prv. 
However, CFI methods typically rely on hardware features and/or computational requirements (e.g., instrumentation, storage for large control flow graphs, and/or shadow stacks~\cite{war_in_mem}) that are prohibitively expensive for MCUs~\cite{cflat}. In addition, the general problem of enumerating valid and invalid control flow paths is often intractable~\cite{aces,aliasing}.

Due to the challenges associated with CFI, Control Flow Attestation (\CFA) was proposed in C-FLAT~\cite{cflat}. 
The key idea in \CFA is to outsource the detection of control flow violations to 
the computationally resourceful \vrf (e.g., a back-end server).
To support this remote verification, \prv builds an authenticated log containing all control flow transfers, i.e., the source and destination of all branching instructions (e.g, \texttt{jumps}, \texttt{returns}, \texttt{calls}, etc.) within the execution of a given operation. This log is obtained by either (1)  instrumenting each branching instruction with additional instructions to securely save branch destinations in protected memory~\cite{cflat,oat,tinycfa}; or (2) using customized hardware to detect branches and save their destinations~\cite{lofat, dessouky2018litehax, zeitouni2017atrium}. The produced ``control flow log (\log)'' is authenticated -- usually MAC-ed or signed by a Root-of-Trust (RoT) in \prv -- 
and sent to \vrf along with an \RA report. 
In possession of both the attested binary and the log of all control flow transfers, \vrf can check if the reported control flow path is valid and is even able to emulate the reported execution (if data inputs are provided, as in~\cite{dialed}).

\subsection{\CFA Limitations: Auditing \& Healing}
Following C-FLAT~\cite{cflat}, additional \CFA designs were presented~\cite{lofat, tinycfa, oat, dessouky2018litehax, zeitouni2017atrium} under various assumptions and guarantees.
Despite substantial progress, current \CFA techniques share several limitations. Due to their passive nature, they offer no guarantee that a \log is ever received by \vrf in case of \prv compromise. 
While this suffices to detect compromises (in general, absence of an \RA report indicates that \underline{something is wrong}), it precludes {\it auditing} \log to pinpoint the source of compromises (i.e., to determine \underline{what is wrong}).
The latter is non-trivial to obtain, since a compromised \prv might ignore the protocol and simply 
refuse to send back reports that indicate a compromise.
Furthermore, current techniques cannot guarantee \prv's remediation when a compromise is detected. 

In addition, current techniques manage \log in ways that introduce non-trivial challenges. \reqchange{5}{A typical approach is to compute an in-order hash-chain of all entries in \log. While this reduces the required storage (only the latest hash value needs to be maintained by \prv), it requires \vrf to have {\it a priori} knowledge of all valid control flow paths. Without this knowledge, \vrf cannot compute the correct hash result during the verification of \CFA report. Similar to the CFI case, determining all valid control paths is non-trivial and often infeasible. An alternative approach is to store \log in its entirety and send it verbatim to \vrf~\cite{oat}, once the attested execution is over. This eases the verification process. However, as \log grows rapidly, it can quickly fill up \prv's limited memory. Due to this limitation, some \CFA techniques (e.g., OAT~\cite{oat} and Tiny-CFA\cite{tinycfa}) are only envisioned for small and self-contained operations.
In SCaRR~\cite{scarr}, this limitation is resolved by requiring \prv to transmit a series of intermediate logs of reduced size, rather than the entire \log.
This allows continuous verification of the program's control flow using a series of fine-grained reports.
However, since SCaRR was designed for high-end cloud systems, its applicability for low-end MCUs remains unclear.}

Finally, existing \CFA architectures either (1) rely on 
code instrumentation, resulting in substantial runtime and binary size overhead; or (2) rely fully on hardware features that are prohibitively expensive to low-end MCUs.

\subsection{Contributions: Efficient Control Flow Auditing \& Active Compromise Remediation}

This paper proposes \acron: an Active Control Flow Attestation architecture. \acron addresses aforementioned limitations by composing concepts from \CFA and Active RoTs (see Section~\ref{sec:garota}) to guarantee that \vrf always receives \log and is able to remotely remediate \prv's state in case of compromise detection. \acron architecture is implemented as an inexpensive and open-source hardware/software co-design.
In sum, \acron anticipated contributions are threefold:
\begin{compactitem}
    \item We propose \acron, the first architecture to guarantee \vrf eventually receives \CFA reports (\log-s) containing \prv's execution trace. 
    \acron also supports guaranteed healing of \prv when a compromise is detected. These features are obtained through a synergic combination of Active RoTs, \CFA, and novel \CFA-specific non-maskable interrupts, realized as a hybrid (HW/SW) design.
    
    \item While prior hybrid approaches exist in \RA, current \CFA techniques relied either on customized (and relatively expensive) hardware or on software instrumentation. 
    We present the first hybrid design for \CFA that eliminates any software instrumentation requirements and minimizes hardware cost, making it affordable even to simple MCUs.
    \acron also demonstrates the feasibility of secure \log slicing (introduced by SCaRR~\cite{scarr}) in MCUs and leverages this feature to support fine-grained control flow auditing of arbitrarily sized software operations.

    \item We propose a continuous \acron protocol aimed at on-demand sensing/actuation use cases. The protocol integrates \acron with a typical on-demand MCU application: MCU awaits for command(s) $\rightarrow$ performs action(s) $\rightarrow$ reports result(s) $\rightarrow$ returns to idle/waiting state. We provide open-source end-to-end implementations and demonstrative videos of such use-cases (including \acron implementation) on an FPGA-based deployment in~\cite{acfarepo}.
\end{compactitem}

%

\section{Background}
\label{sec:background}

\subsection{Scope}
\label{sec:scope}

\reqchange{4}{This work focuses on simple MCUs and aims for minimality of hardware requirements.
We argue that a design that is cost-effective enough for
the lowest-end MCUs could also be adapted and potentially enriched for higher-end devices, with less strict hardware budgets (we discuss alternative designs in Section~\ref{sec:alternatives}). Adapting designs in the other direction is more challenging. The choice of a simple device also facilitates reasoning and presenting \acron concepts systematically.}

Following these premises, we present a design for low-end MCUs based on low-power single-core platforms with only a few kilobytes (KB) of program and data memory (such as Atmel AVR ATmega and TI MSP430). They feature $8$- and $16$-bit CPUs, typically running at $1$-$16$ MHz clock frequencies, with $\approx64$ KB of addressable memory. SRAM is used as data memory ($DMEM$) ranging in size between $4$ and $16$ KB, while the rest of the address space is available for program memory ($PMEM$). They run software at ``bare metal'', executing instructions in place (physically from $PMEM$), and have no memory management unit (MMU) to support virtual memory.

\acron prototype is implemented atop a representative of this class of devices: the well-known TI MSP430 ultra low-energy MCU.
This choice is simply due to the availability of an open-source version of the MSP430 hardware from OpenCores \cite{openmsp430}. 
\reqchange{4}{Nevertheless, we expect \acron design to generalize to other bare-metal MCUs (e.g., ARM Cortex-M). See Section~\ref{sec:alternatives} for future work discussion on adapting \acron to higher-end devices (e.g., those featuring virtual memory).}

\subsection{Remote Attestation (\RA)}\label{sec:bg_ra}


\RA is a challenge-response protocol between \vrf and \prv.
It allows \vrf to remotely assess \prv's trustworthiness by measuring the content of \prv's memory.
As depicted in Figure~\ref{fig:RA}, a typical \RA interaction involves the following steps:
\begin{figure}[th]
    \centering
    \includegraphics[width=.5\columnwidth]{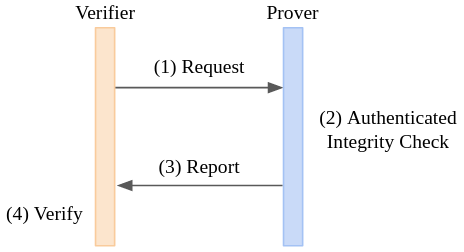}
    \caption{A typical \RA interaction}
    \label{fig:RA}
\end{figure}
\begin{compactenum}
    \item \vrf requests \RA from \prv by sending a cryptographic challenge $Chal$.
    \item Upon receiving $Chal$, \prv computes an authenticated integrity-ensuring function over its own memory and $Chal$, producing report $H$.
    \item \prv sends the report $H$ back to \vrf.
    \item \vrf checks $H$ against an expected value to determine if \prv has been compromised.
\end{compactenum}

The authenticated integrity-ensuring function in step 2 is implemented using a message authentication code (MAC) or a digital signature. The secret key used in this operation must be securely stored to ensure that it is inaccessible to any untrusted software on \prv. \RA threat models (including the one considered in this paper) assume that \prv is susceptible to full software compromise. Therefore, secure storage for the \RA secret key implies some level of hardware support.

\RA architectures are generally classified in three types: 
software-based (a.k.a. ``keyless''), hardware-based, or hybrid. Software-based architectures~\cite{KeJa03, SPD+04, SLS+05, SLP08} require no hardware support. However, \RA must be local (e.g., over a one-hop wired communication) and requires several strong assumptions about \adv capabilities, implementation optimality, and fixed communication delays that are often infeasible in practice~\cite{SW-Attsnt-Attacks}. Hardware-based architectures~\cite{PFM+04, KKW+12, SWP08} rely on standalone cryptographic coprocessors (e.g., TPMs~\cite{tpm}) or complex support from the CPU instruction set architecture (e.g., Intel SGX~\cite{sgx}). Although these approaches provide strong security guarantees for \RA, their hardware cost is often too expensive and unrealistic for MCUs. Hybrid architectures~\cite{vrased, smart, tytan} focus on low-cost MCUs. They leverage minimal hardware support to store cryptographic secret(s) and to support secure execution of a software implementation of the integrity-ensuring function (MAC or signature) computed during the \RA protocol. Hybrid architectures aim to combine the low hardware cost of software-based approaches with the security guarantees offered by hardware-based approaches.

VRASED~\cite{vrased} is a formally verified hybrid \RA architecture.
It implements the authenticated integrity-ensuring function in software while introducing small trusted hardware to enforce the correct execution of this software and confidentiality of the \RA secret key. In addition, VRASED guarantees that the attested memory is temporally consistent, i.e., not modifiable during the memory measurement. We further elaborate on \RA related work in Section~\ref{sec:relatedwork}.
As we discuss in Sections~\ref{sec:design_overview} and~\ref{sec:design}, \acron hybrid design leverages VRASED to replace relatively costly hardware-based hash engines and to authenticate \CFA reports.

VRASED also provides an optional design extension that supports authentication of \vrf requests. In this case, an authentication token accompanies \vrf requests.
\vrf computes this token as a MAC over \chal, using the \RA key. 
To prevent replays, \chal must be a monotonically increasing counter, and the latest \chal used to successfully authenticate \vrf must be stored in \prv's persistent and protected memory. In each \RA request, incoming \chal must be greater than the stored value. Once an \RA request is successfully authenticated, the stored value is updated accordingly. As discussed later in Section~\ref{sec:design}, \acron also uses this \vrased extension to authenticate remediation decisions made by \vrf.

\subsection{Control Flow Attestation (\CFA)}

In addition to the \RA result, \CFA also provides \vrf with an authenticated \log that contains the order in which the instructions in the attested binary were executed.
\log is either produced by dedicated hardware or obtained by instrumenting each branching instruction with additional instructions to securely save their source and destination addresses in protected memory. Once the execution of the attested operation completes, \log is authenticated (usually MAC-ed or signed by the \RA RoT in \prv) and reported to \vrf along with the \RA result. In possession of both the attested binary and \log, \vrf can decide if the reported control flow path is valid, and thus if \prv has been compromised.

Prior \CFA designs (and \RA/\PoX architectures, more broadly) consider absence of a valid report (step (3) in Figure~\ref{fig:RA}) as a sign that \prv is compromised, as the honest \prv would have followed the protocol.
Such an assumption is sensible from a {\it detection} perspective. 
However, it prevents \vrf  from {\it securely auditing} the source of an exploit -- the control flow violation leading to the exploit may never be received by \vrf; thus \vrf cannot easily pinpoint the vulnerability. One of \acron core contributions is to enable {\it secure runtime auditing}, i.e., guaranteeing delivery of \log, even when \prv is compromised by malware that prevents \prv from sending \CFA reports to \vrf. \reqchange{2}{Naturally, this guarantee holds when \adv is unable to jam the network indefinitely. In the case where \adv has such capability and \vrf never receives reports, \acron can optionally halt execution on \prv.}

C-FLAT~\cite{cflat} was the first proposal for \CFA. It uses ARM TrustZone's secure world as an RoT to build and store \log. In a pre-processing phase, a control flow graph (CFG) is constructed and each node in the CFG is assigned a unique Node ID. The executable is instrumented with secure-monitor calls to TrustZone secure world to save Node IDs whenever a node transition occurs. C-FLAT measurement engine, implemented within the secure world, extends a hash-chain with the Node ID on each call. Once execution of the attested task completes, the hash-chain uniquely identifies the control flow path.
%
Subsequent \CFA architectures~\cite{lofat,tinycfa,oat,dessouky2018litehax,zeitouni2017atrium,scarr,recfa} built upon 
C-FLAT (see Section~\ref{sec:relatedwork}). Despite substantial progress, to the best of our knowledge, the problems identified in Section~\ref{sec:intro} remain common to all of these \CFA architectures.

\subsection{Active RoTs}\label{sec:garota}

Classic attestation methods (including \RA and \CFA) have been designed as passive RoTs. 
As such, they can detect compromises to \prv integrity.
However, they cannot guarantee actions will be taken beyond detection.
Recently proposed \emph{active} RoTs~\cite{garota,proactive1,proactive2,proactive3,aion}, on the other hand, focus on availability under software compromise.
In particular, GAROTA~\cite{garota} is a generalized interrupt-based active RoT designed as a hardware monitor for low-end MCUs. It supports a secure association between a \trigger event (e.g., ``temperature exceeds a threshold'') and the correct execution of a software function responsible for a safety-critical action (e.g., ``sound the alarm''), whenever the \trigger event occurs. This guaranteed \trigger-action association must hold even when the MCU software is compromised.

To achieve this goal, GAROTA provides two core features: \textit{guaranteed triggering} and \textit{re-triggering on failure}. \textit{Guaranteed triggering} ensures that a predefined trusted software function (\func) always takes over execution when a corresponding safety-critical interrupt of interest -- the \trigger~-- occurs. After the \trigger, \func execution cannot be tampered with or interrupted until its completion (i.e., reaching its pre-defined exit instruction). Any attempt to interfere with \func execution causes an immediate MCU reset.
The reset brings the MCU back to a clean state where interrupts and Direct Memory Access (DMA) controllers are disabled. Immediately after any reset, the \textit{re-triggering on failure} property ensures that \func is always the first software to execute. Therefore, malware on \prv is unable to prevent \func from executing in its entirety once a \trigger has occurred. At best, malware can cause a reset by attempting to interrupt \func. The reset will, in turn, lead to a secure re-execution of \func with interrupts and DMA disabled. 

GAROTA is a general architecture that supports any pre-existent interrupt source to be configured as a \trigger, including GPIO inputs (i.e., external inputs from sensors, buttons, etc.), timers, and (UART-based) network events. As \func is a software function, it can implement any desired action that should take place following the \trigger event.
To obtain this secure \trigger-action association, GAROTA hardware monitors execution and protects the initial configuration of the \trigger interrupt from illegal modifications or disablement. This protection includes preserving interrupt configuration registers, interrupt handlers, and the interrupt vector table. This way, GAROTA guarantees that a trigger always results in an invocation of \func.
However, guaranteed invocation of \func upon occurrence of a \trigger is not sufficient to claim that \func is properly performed, since the \func code (and execution thereof) could be tampered with. To this end, GAROTA hardware also provides runtime protections that prevent any unprivileged/untrusted program from modifying \func code. GAROTA monitors the execution of \func to ensure:
\begin{compactenum}
\item {\bf Atomicity:} \func executes uninterrupted, from its first instruction (legal entry), to its last instruction (legal exit);
\item {\bf Non-malleability:} $PMEM$ region storing \func implementation is unmodifiable at runtime. During \func execution, $DMEM$ can only be modified by \func itself, e.g., no modifications by DMA controllers.
\end{compactenum}
These properties ensure that any malware potentially residing on the MCU (i.e., compromised software outside \func or compromised DMA controllers) cannot tamper with \func execution.

\acron builds atop active RoT concepts as one of its features to guarantee secure control flow auditing and device healing. Unlike \garota, \acron creates a new \trigger, based on \CFA-specific events, implemented as a non-maskable interrupt that is controlled only by \acron, in hardware.
The associated \func in \acron implements a sequence of actions to guarantee that \log is always received by \vrf and that a \vrf-initiated remediation function is properly invoked, when applicable.

\section{\acron High Level Overview}
\label{sec:design_overview}

This section presents \acron high level ideas, before going into its details in Section~\ref{sec:design}. 
To construct \log, \acron implements a hardware \CFA monitor that detects and saves all control flow transfers that happen during the attested execution to a fixed dedicated $DMEM$ region. The monitor also ensures that this region is read-only to all software. Therefore, compromised \prv software is unable to modify \log.
When reporting the \CFA result (including both \prv binary and \log) to \vrf, \acron offers the following key features:
\begin{compactenum}
 \item[\textbf{[F1]}] {\bf Secure Control Flow Auditing:} it guarantees that any \log (or partial \log, when \log is sliced and streamed due to limited storage) 
 generated by the \CFA hardware monitor must be received, successfully authenticated, and accepted by \vrf. \reqchange{2}{The active \CFA RoT in \prv assures that execution remains paused until a confirmation of receipt from \vrf reaches \prv. In the interim, the report can be periodically re-transmitted to \vrf to cope with occasional network losses. Optionally, if an (application-specific) upper bound on the wait time is reached without receiving \vrf confirmation, \prv can automatically switch to the remediation phase (see below) or resume execution, depending on the desired policy (strict vs. best-effort). Our discussion focuses on a strict version, where software integrity is more important than minimizing disruption. In this case, \prv must always wait for \vrf confirmation. Thus, \prv execution halts if messages are discarded indefinitely. We revisit alternative designs for the best-effort case in Section~\ref{sec:alternatives}.}
 \item[\textbf{[F2]}] {\bf Guaranteed Remediation:} as a part of its confirmation, \vrf can indicate whether \prv execution is allowed to proceed normally, i.e., when the \CFA verification indicates a benign and expected state. In case of compromise detection, \vrf can indicate that \prv must switch to the remediation phase. 
 \acron ensures that \vrf command is processed, irrespective of a compromised software state on \prv. The specific remediation action is configurable, depending on the desired policy for each particular application domain. For instance, it might include remotely updating the binary in $PMEM$, erasing all memory, or shutting \prv down.
\end{compactenum}

 At its core, \acron implements an Active RoT (recall Section~\ref{sec:garota}) with a \trigger used to take over \prv execution whenever \log (or a slice of \log, if \log sliced and streamed) 
 must be sent to \vrf. To that end, \acron hardware monitor implements a new secure interrupt occurring in three cases (whichever comes first):

 \begin{compactenum}
    \item[\textbf{[T1]}] when a timer expires, imposing periodic reports to \vrf;
    \item[\textbf{[T2]}] when the \log designated memory is full, implying that its contents must be received by \vrf and flushed before new control flow transfers can be stored;
    \item[\textbf{[T3]}] when \prv resets/boots or when the attested operation concludes its execution.
\end{compactenum}

 We note that \trigger case~{\bf [T2]} implies that whenever \prv runs out of dedicated memory to store \log, the partial snapshot of the control flow transfers in \log is automatically authenticated and transmitted to \vrf for verification. After this step, the same memory region can be re-used to store subsequent control flow transfers.
 
 In \acron, the associated \trigger-handling function \func is referred to as Trusted Computing Base (\tcb) Software. It implements three steps within itself:
 \begin{compactitem}
   \item {\bf \tcb-Att:} is an \RA RoT implemented using VRASED and is always called upon \trigger to measure (i.e., compute a MAC of) \prv binary and the current \log;
   \item {\bf \tcb-Wait:} always follows \tcb-Att and is called to send the report (computed by \tcb-Att) to \vrf and wait for \vrf decision on whether a remediation phase should follow (in case of compromise detection);
   \item {\bf \tcb-Heal:} implements the remediation action that may occur based on \vrf decision after analyzing the report.
 \end{compactitem}

\begin{figure}[t]
    \centering
    \includegraphics[width=\columnwidth]{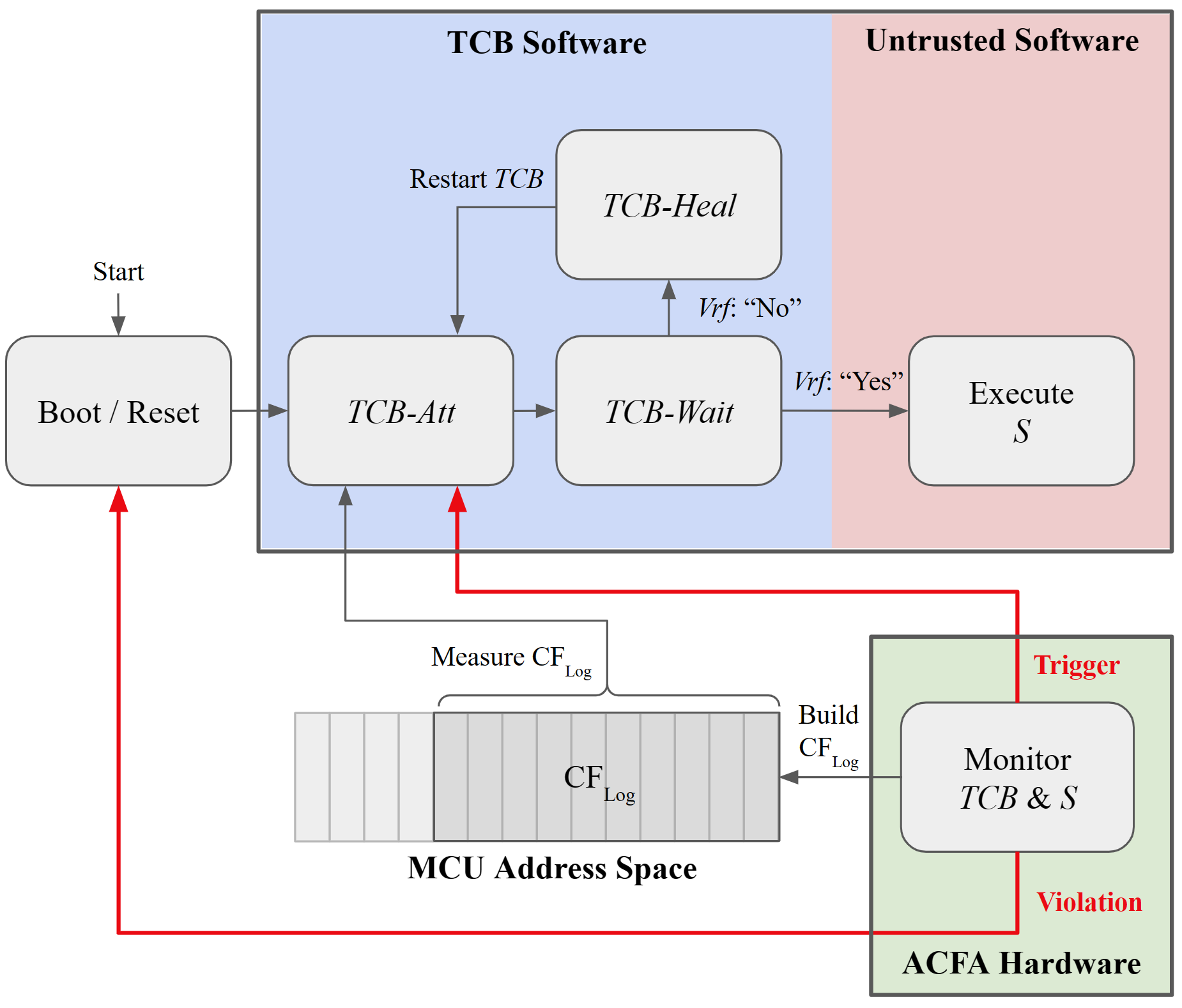}
    \caption{\acron Execution Workflow}
    \label{fig:phases}
\end{figure}

Figure~\ref{fig:phases} illustrates \acron execution workflow alongside \acron HW module responsible for (1) generating and protecting \log; and (2) issuing the \trigger interrupt when a \CFA report must be sent to \vrf. 
%
%
%
\reqchange{2}{We highlight two important consequences of \acron design and execution workflow. Even compromised software on \prv is unable to preclude sending of \log to \vrf, as \trigger cannot be disabled due to the active RoT guarantees and the sending function is implemented within the (atomically executed) \tcb. Therefore, \vrf receives \log even in case of \prv compromise, 
enabling auditing of the exploit's control flow path to identify the vulnerability source. Similarly, \tcb-Heal is also implemented within \tcb and cannot be avoided by any external attempts originating from potentially compromised software on \prv.}

With this design, we envision \acron to be particularly useful in security-critical on-demand sensing applications, where \prv is expected to perform sensor readings upon receiving \vrf commands.
An end-to-end \acron implementation with a sample application is presented in Section~\ref{sec:end_to_end}.

\begin{figure}[t]
    \centering
    \includegraphics[width=0.7\columnwidth]{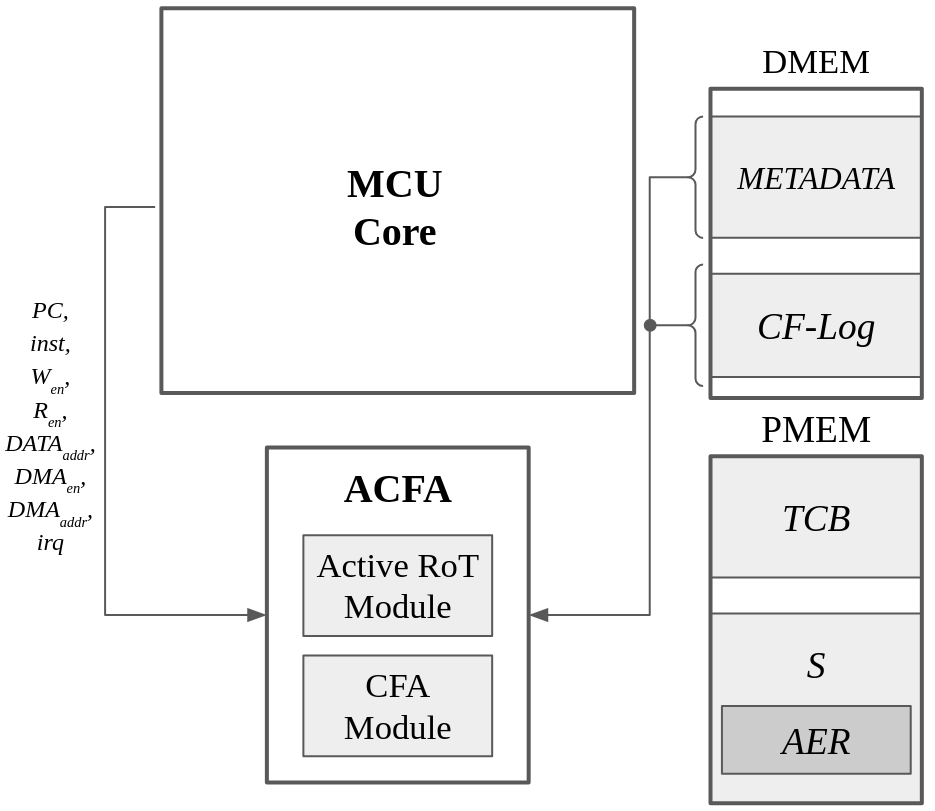}
    \caption{\acron Architecture}
    \label{fig:system}
\end{figure}


\section{\acron in Detail}
\label{sec:design}

This section presents \acron details. We start by defining the adversary model, \acron architecture, and \acron protocol. We then deconstruct \acron design into multiple required security properties and associated design elements that enforce each required property. Finally, we analyze the security of the overall construction according to the adversary model.

\subsection{Adversary (\adv) Model}\label{sec:adv_model}
 
We consider that \adv can exploit software vulnerabilities in \prv software to (1) modify any writable memory that is not explicitly protected by hardware-enforced access controls; (2) cause malicious control flow transfers on untrusted software; and (3) attempt to hide their malicious actions (in the form of injected code or hijacked control flows).
Unless prevented, modifications to program memory can change instructions, and modifications to data memory can corrupt intermediate computation results, affecting the program's intended control flow. 
\adv may also attempt to trigger interrupts or re-program any interrupt handler to achieve similar goals; re-programming an interrupt handler can be done by either modifying its software directly or modifying an entry in the interrupt vector table to point to any other (potentially malicious) software.
In addition, \adv has a Dolev-Yao~\cite{dolev-yao} capability with respect to the network. Therefore, it may discard, inject, or attempt to modify messages between \prv and \vrf.
Hardware attacks that require physical access to circumvent \prv hardware protections (or hardware-protected software) are out-of-scope in this paper. Protection against the latter involves orthogonal physical access control measures~\cite{ravi2004tamper,obermaier2018past}.

\reqchange{2}{\textit{\textbf{Remark: }{As noted in Section~\ref{sec:design_overview}, \acron aims to guarantee \log delivery and \prv remediation assuming eventual communication. In case of a Dolev-Yao \adv that discards all messages indefinitely, \acron (in its strict version) intentionally halts \prv's compromised execution.}}}

\subsection{\acron Architecture}

Figure~\ref{fig:system} presents \acron architecture. \acron HW interacts with the MCU Core and with main memory. It is composed of two sub-modules: the Active RoT Module and the \CFA Module.
 
\begin{table}[t]
    \caption{Notation Summary}
    \label{tab:symb}
    \resizebox{\columnwidth}{!}{%
    \centering
    \begin{tabular}{|c|c|}
        \hline
        \textbf{Symbol} & \textbf{Definition} \\
        \hline
        PC & Program Counter (points to the instruction currently being executed). \\
        \hline
        \inst & Bits of the currently executing instruction that specify its operation type \\
        \hline
        $W_{en}$ & MCU write enable bit (is set whenever the CPU writes to memory)\\
        \hline
        $R_{en}$ & MCU read enable bit (is set whenever the CPU reads from memory)\\
        \hline
        $D_{addr}$ & MCU data address signal (contains the that address being read -- when $R_{en}$ is set\\ 
                   & -- or written -- when $W_{en}$ is set -- at each execution cycle). \\
        \hline
        $DMA_{en}$ & DMA enable bit (is set whenever the DMA reads or writes from/to memory)\\
        \hline
        $DMA_{addr}$ & DMA data address signal MCU data address signal (contains the that address\\
                     & being read or written when $DMA_{en}$ is set is set -- at each execution cycle)\\
        \hline
        $irq$ & MCU signal that is set when an interrupt is occurring\\
        \hline
        \textit{\tcb} & Trusted computing base (PMEM location storing \acron trusted software)\\
        \hline
        \textit{S} & $PMEM$, except for \tcb, i.e., region storing all untrusted application code\\
        \hline
        \er & Region storing code whose execution is to be attested. Located within $S$.\\
        \hline
        \textit{METADATA} & \acron-reserved $DMEM$ region used to store \acron-associated data\\
        \hline
        \log & Log that stores control flow transfers during \er execution\\
        \hline
        $\ptr$ & Current size of \log\\
        \hline
        \end{tabular}
    }
\end{table}

The MCU address space consists of program memory ($PMEM$) and data memory ($DMEM$). 
In \acron, $PMEM$ is divided between the \tcb software and other (untrusted) application software, denoted \s. \tcb is located in a fixed memory region. \acron protects this region by checking MCU signals at runtime. 
The Attested Executable Region (\er) is a subset of \s containing the software of interest that should be attested/audited by \vrf. 
\er location and size are configurable. Therefore, \vrf can define what should be attested/audited: the execution of a code segment, a single function, multiple functions, or the entirety of $PMEM$.
\acron also reserves regions $METADATA$ and \log in fixed physical locations of $DMEM$.
$METADATA$ is used to store \acron-related variables: the cryptographic challenge $Chal$ (received from \vrf), addresses defining the the boundaries of $\er$ in memory ($\er_{min}$, $\er_{max}$), and the current size of \log ($\ptr$). 

\acron hardware 
monitors 
several MCU signals in order to enforce security properties. Table~\ref{tab:symb} summarizes the notation used in the rest of this paper, including CPU signals monitored by \acron HW.
Among these signals, the program counter ($PC$) contains the address of the current instruction being executed. This signal tells \acron HW which software region (\tcb, \s, or $\er$) is executing. The \inst signal contains the ``opcode'' of the currently  executing instruction (as a bit-string). \inst is used by \acron to determine if a branch instruction is occurring. \acron also monitors signals related to memory accesses -- the write and read enable bits ($W_{en}$,$R_{en}$) and the data address ($D_{addr}$) being accessed by the MCU. This allows \acron to determine if a read/write is occurring and the respective memory address of the read/write operation, enabling prevention of illegal reads/writes.
Similarly, $DMA$ access signals ($DMA_{en}, DMA_{addr}$) are also monitored to detect $DMA$ reads/writes and their destinations. Finally, \acron also monitors signals related to interrupts such as the interrupt bit ($irq$) and the global interrupt enable bit ($gie$) to detect when interrupts are triggered, accepted, and enabled.

\acron HW is composed of two sub-modules: the Active RoT module and the \CFA module. Based on the aforementioned HW signals, they enforce several required security properties that will be described in detail in Section~\ref{sec:submodules}.

The Active RoT Module is responsible for the {\it guaranteed triggering} and {\it re-triggering on failure} properties (see Section~\ref{sec:garota}) which guarantee the correct execution of \acron TCB Software.
\acron protects each \tcb-\trigger source (\textbf{[T1]-[T3]}) from malicious software by implementing them as Non-Maskable Interrupts (NMIs). As opposed to normal interrupts, NMIs cannot be disabled in software.
\acron also ensures that the deadline of the periodic timer (used by trigger \textbf{[T1]}) is only configurable by TCB Software (i.e., when $PC \in TCB$). Similarly, \acron creates a new NMI that is triggered whenever \log region is full (\textbf{[T2]}) or
when \er execution is concluded (\textbf{[T3]}) by triggering the NMI when $PC = \er_{max}$.

To assure integrity of the \CFA report, the \CFA Module 
detects any illegal attempts to modify data associated with the execution of \er (such as $METADATA$, \log, and \er binary itself), as this data is included in the \CFA report and used by \vrf to interpret such report. \CFA Module also detects and logs all control flow transfers (due to branches or interrupts) onto \log in an optimized fashion.


Any violation to \acron properties (as detected by \acron HW) triggers an MCU reset (recall Figure~\ref{fig:phases}). A reset implies execution of \tcb. Therefore, \acron ensures that an exploit always 
leads to \vrf receiving a \CFA report that contains the exploit's control flow information, allowing \vrf to pinpoint the source of this exploit.


%

\reqchange{3}{\textit{\textbf{Remark:} As shown in Figure~\ref{fig:system}, \acron operates in parallel with the MCU Core's execution pipeline. Hence, the execution critical path delay is not affected by \acron.}}

\subsection{\acron Protocol}\label{sec:protocol}

\input{protocol}

Figure~\ref{fig:prot} presents \acron protocol.
A protocol instance starts when \tcb is invoked on \prv due to one of \acron triggers ({\bf {[T1], [T2], or [T3]}}).
Recall that \acron triggers include, boot/program end, expiration of a timer, and \log  being full.
The timeout parameter can be configured to meet application needs. For instance, to minimize disruption in case of on-demand sensing applications, the deadline can be set to give sufficient time for the sensing code to complete its execution while still assuring that the report is always received by \vrf in a timely manner. 

\acron protocol implements the execution workflow illustrated in Figure~\ref{fig:phases}. 
Once \tcb is invoked in \prv, it executes \tcb-Att to produce an \RA measurement $H$, as in Step 1 of Figure~\ref{fig:prot}. 
$H$ is computed on $PMEM$, $METADATA$ and \log, using a key ($\attkey$) that is pre-shared between \vrf and the RoT in \prv.
Then, in Step 2, \prv transitions to \tcb-Wait, generating and sending an \acron report to \vrf. This report consists of $H$, $METADATA$ and \log. It then awaits for \vrf response.
Upon receiving \acron report, \vrf performs the verification process (\vrfy), in Step 4, by:
\begin{compactenum}
    \item Checking validity of $H$. As \vrf possesses \prv expected binary (denoted $PMEM'$), this check can be done by computing the expected $H$, i.e: 

    \begin{center}
    $H \stackrel{?}{=} \cfattest_\attkey(PMEM', METADATA, \log)$
    \end{center}
    \item Checking if $METADATA$ matches \chal and \er boundary as requested by \vrf in the previous instance of \acron protocol. We note that when the protocol runs for the first time, \prv has yet to receive any challenge or \er boundary from \vrf. In this case, \vrf instead compares $METADATA$ with a default value, i.e., $0$.
    \item Evaluating \prv reported execution trace based on \log and its size ($\ptr$), where $\ptr$ is located inside $METADATA$. 
    This step can employ a variety of techniques, such as evaluating \log on \er control flow graph or emulating a shadow stack for \er execution. We discuss instantiations of \vrf in Section~\ref{sec:cfa-ver}.
\end{compactenum}

If verification succeeds, \vrf approves the report and thus sets the approval flag ($app:=1$), indicating that \prv is allowed to continue execution; otherwise, the approval flag is cleared ($app:=0$).
In Step 5, \vrf creates an \acron response by incrementing the challenge $\chal'$, defining contiguous region of PMEM [$\er_{min}$, $\er_{max}$] that determines the next operation to be audited (which could remain the same), and computing an authentication token \auth.
This response is forwarded to \prv in Step 6.

Upon receiving the response, \prv authenticates \vrf message in Step 7 (including whether $\chal'$ > \chal, for freshness).
If authentication fails, \prv goes back to waiting in Step 2.
Only when authenticity of the response is confirmed, \prv determines whether \vrf approves the report. 
In case that the report is not approved ($app=0$ in the response), \prv enters \tcb-Heal to perform a remediation operation (e.g., system reset, software update) in Step 8. After the remediation finishes, it restarts the whole process from Step 1 in order to convince \vrf that the remediation was indeed performed successfully by attesting the new system state.

When \vrf approves ($app=1$), \prv 
is authorized to exit \tcb and continue to Step 9 where \prv begins executing the sensor application or resumes execution
from where it left off before the \tcb-\trigger. While executing $\er$, \acron hardware monitors execution and constructs \log. This continues until the occurrence of a new \trigger, which in turn initiates a new instance of the \acron protocol from Step 1.

To support the computation of $H$ as well as the authentication of \vrf message in Step 7, \acron leverages \vrased for \RA (recall VRASED description from Section~\ref{sec:bg_ra}) which ensures that the secret key \attkey used for \RA and for authentication of \vrf message is not leaked, even in case of a compromised software state on \prv. We also note that, to deal with network failures, \acron report and response messages can be re-transmitted periodically, if the subsequent message in the protocol is not received from the respective communication end-point after a given time.
\subsection{Required Security Properties}
\label{sec:prop}

To support the correct execution of \acron protocol defined in Section~\ref{sec:protocol}, irrespective of a potentially compromised software state in \prv, \acron enforces multiple properties to assure \log Integrity (Properties \textbf{[P1-P3]}) and \textit{\tcb} Execution Integrity (Properties \textbf{[P4-P5]}).

\vspace{1mm}
\noindent \textbf{[P1] Read-Only \log:}
\log is read-only to \textit{all} software.
This property is necessary to ensure \log integrity, i.e., \adv cannot tamper with the content in \log.
Without this property, \adv could forge a valid control flow log without executing the intended software by simply overwriting the \log region.

\vspace{1mm}
\noindent \textbf{[P2] \textit{METADATA} Integrity:}
The \vrfy algorithm (Step 4, in Figure~\ref{fig:prot}) depends on \textit{METADATA}. For this reason, \acron guarantees \textit{METADATA} can only be overwritten by \tcb Software, which sets $METADATA$ according to \acron response (sent by \vrf in Step 6 of Figure~\ref{fig:prot}).
$METADATA$ stores the bounds ($\er_{min}$, $\er_{max}$), defining \textit{\er} region. \acron detects and logs control flow transfers based on these boundaries. In addition, the $METADATA$ contains the current size of the log ($\ptr$) and the cryptographic challenge ($Chal$). $\ptr$ determines 
the total
control flow transfers that happened since the last \acron response message (as \trigger may occur before \log is full, due to a timer expiration or a violation of \acron rules). $Chal$ assures that subsequent $ACFA$ reports cannot be replayed.
Thus, $METADATA$ must be write-protected from \prv untrusted software. 

\vspace{1mm}
\noindent \textbf{[P3] \log Correctness:}
All control flow transfers within \er (including any external jump into $\er$, e.g., to invoke \er) must be correctly detected and accurately recorded to \log. 
In other words, \log must reflect the exact sequence of control flow transfers that have happened during \er latest execution (since receipt of the latest \chal).

\vspace{1mm}
\noindent \textbf{[P4] Guaranteed \tcb Triggering/Re-Triggering:}
\trigger must result in guaranteed \tcb execution upon occurrence of {\bf[T1]}, {\bf[T2]}, and {\bf[T3]} cases (defined in Section~\ref{sec:design_overview}). 
In case of {\bf [T1]}, \tcb must be triggered periodically, to enable auditing of time sensitive tasks. The period is configurable from within \tcb.
\tcb must also be triggered when \log is full \textbf{[T2]}, to free \log for new control flow transfers.
Lastly, {\bf [T3]} requires \tcb execution to be triggered on a reset/boot or when \er execution completes.
The former is necessary to prevent \adv from triggering resets to avoid auditing of \acron report by \vrf or to avoid the active remediation phase. 
The latter is required 
since the reaching of $\er_{max}$ may occur before \log is full and before the expiration of the timer.

\vspace{1mm}
\noindent \textbf{[P5] \tcb Integrity:}
Since \tcb is responsible for various security-critical operations in \acron, its integrity is crucial. Its instructions must be write-protected from all other software in \prv. Once called, \tcb must execute atomically, i.e., it faithfully executes the sequence \tcb-Att $\rightarrow$ \tcb-Wait $\rightarrow$ \tcb-Heal, without interruptions and without interference from other software in \prv. 
Atomicity should also prevent jumps to the middle of \tcb, as they could be used to initiate out-of-order execution of \tcb code.
The \RA result in \tcb-Att (Step 1 in Figure~\ref{fig:prot}) must be unforgeable to assure that \acron report is authentic (in turn, this requires absolute confidentiality of the \RA secret key). 
Similarly, \acron response sent by \vrf (Step 6 in Figure~\ref{fig:prot}) defines actions to be taken on \prv and therefore must be authenticated by \tcb. The \RA RoT in \acron must implement these functions securely, irrespective of any compromised software outside \tcb.

\subsection{Specification of \acron Components}
\label{sec:submodules}

\begin{figure}
\scriptsize
\fbox{
    \parbox{0.95\columnwidth}{
        \textbf{\underline{\textit{Machine Model:}}}\\

        1. Memory Modification:
        \begin{equation*}
        \begin{split}
        \mathit{modifyMem(X)} \equiv
        \mathit{(W_{en} \land D_{addr} \in X) \lor (DMA_{en} \land DMA_{addr} \in X)}
        \end{split}
        \end{equation*}

        2. \inst signal contains opcode of the instruction pointed by $PC$\\

        3. $\mathit{call_{irq}}$ bit is set whenever a control flow transfer occurs due to interrupt handling. 
    }
}
\caption{Machine Model}
\label{fig:machine}
\end{figure}

We now discuss how \acron enforces~\textbf{[P1]-[P5]}, presented in Section~\ref{sec:prop}. In particular, we define the logic required to implement these properties based on the MCU signals monitored by \acron HW (recall \acron monitored signals from Table~\ref{tab:symb}). This logic is then implemented by \acron HW. 

\begin{figure}
    \centering
    \includegraphics[width=0.9\columnwidth]{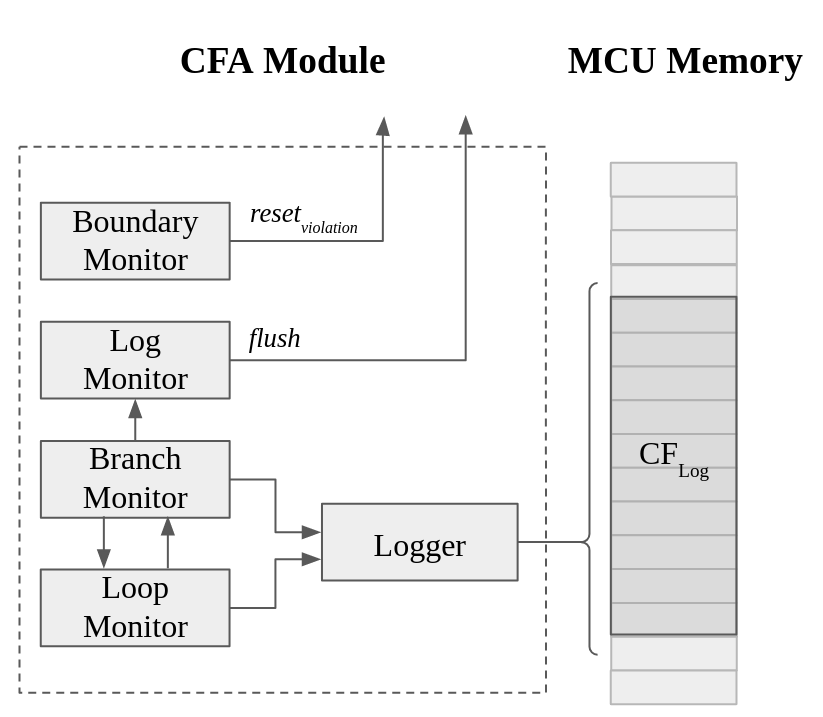}
    \caption{Sub-Modules within the \CFA module}
    \label{fig:hw}
\end{figure}


Figure~\ref{fig:machine} defines the MCU machine model based on how the MCU behavior is reflected in its hardware signals. First, $\mathit{modifyMem(X)}$ predicate models MCU signals whenever a given memory address $\mathit{X}$ is modified by either CPU or DMA. In order for the CPU to modify memory region $\mathit{X}$, the $\mathit{W_{en}}$ bit must be set and $\mathit{D_{addr}}$ must point to a location within $\mathit{X}$. Similarly, in order for DMA to modify $\mathit{X}$, $\mathit{DMA_{en}}$ must be set and $\mathit{DMA_{addr}}$ must be within $\mathit{X}$.
The \inst signal contains the opcode determining the instruction that is currently being pointed by $PC$ in $PMEM$. Each distinct instruction in the CPU instruction set architecture has a different opcode.
Finally, $\mathit{call_{irq}}$ bit is set whenever a control flow transfer happens due to interrupt handling.\\


\vspace{-0.6em}
\begin{center}
 {\bf \CFA Module}
\end{center}
\vspace{-0.6em}

The \CFA module in \acron is depicted in Figure~\ref{fig:hw}. It consists of five sub-modules.

$\bullet$ \textbf{Boundary Monitor:} Boundary Monitor enforces \textbf{[P1]} and \textbf{[P2]} as specified in the logic of Figure~\ref{fig:boundary}, preventing unauthorized modifications to \textit{METADATA} and \log. If \textit{METADATA} is modified by any software other than \tcb or if there is a software-write attempt to \log, it resets \prv. 

$\bullet$ \textbf{Branch Monitor: } As specified in Figure~\ref{fig:branchdetect}, Branch Monitor detects control flow transfers by checking the \inst signal to identify branching instructions, i.e., \texttt{call}, \texttt{jmp}, \texttt{ret}, and \texttt{reti}. Additionally, it monitors the $\mathit{call_{irq}}$ signal to detect branches due to interrupts. It outputs $\mathit{branch_{detect}} = 1$ if a control flow transfer is detected. Depending on the CPU state, a variable number of instructions may execute in between the moment when an interrupt is received (when $irq=1$) and the moment when the jump to the ISR is triggered (i.e., the moment when the control flow transfer occurs). Therefore, $\mathit{call_{irq}}$ itself is determined based on a combination of the MCU signals $irq$ and $gie$. We defer these implementation-specific details to Appendix~\ref{ap:callirq}. 

\begin{figure}[t]
\footnotesize
\fbox{
    \parbox{0.95\columnwidth}{
        \underline{\textbf{HW Specification:}} Monitor Boundaries of \textit{METADATA} and \log
        \begin{equation}
        \tiny
        \begin{split}
          &[modifyMem(METADATA) \land PC \notin \tcb] \lor modifyMem(\log) \\& \lor (DMA_{en} \land DMA_{addr} \in METADATA)   \rightarrow reset
        \end{split}
        \end{equation}
    }
}
\caption{Hardware Spec.: Boundary Monitor Sub-Module}
\label{fig:boundary}
\end{figure}

\begin{figure}[t]
\footnotesize
\fbox{
    \parbox{0.95\columnwidth}{
        \textbf{\underline{HW Specification:}} Detect branch due to instruction or interrupt
        \begin{equation}
        \tiny
        \begin{split}
          &\mathit{[(\inst = \mathit{call})\lor(\inst = \mathit{jmp})} \lor (\inst = \mathit{ret}) \lor (\inst = \mathit{reti}) \lor call_{irq}] \rightarrow branch_{detect}
        \end{split}
        \end{equation}
    }
}
\caption{Hardware Spec.: Branch Monitor Sub-Module}

\label{fig:branchdetect}
\end{figure}

\begin{figure}[t]
\footnotesize
\fbox{
    \parbox{0.95\columnwidth}{
        \textbf{\underline{HW Specification:}} Secure Logging
        \tiny
        {\smaller
        \begin{equation}
           \mathit{\ptr = maxSize(\log) \rightarrow log_{full}}
        \end{equation}
        \begin{equation}
            \mathit{(PC \in \er) \land branch_{detect} \land \neg log_{full} \rightarrow hw_{en}}
        \end{equation}
        \begin{equation}
            \mathit{\neg(PC \in \er) \land (PC_{next} \in \er) \land branch_{detect} \land \neg log_{full} \rightarrow hw_{en}}
        \end{equation}
        \begin{equation}
            \mathit{hw_{en} \land \neg log_{full} \land \neg loop_{detect} \rightarrow \ptr\mathtt{++}}
        \end{equation}
        \begin{equation}
            \mathit{PC = \tcb_{max} \rightarrow \ptr = 0}
        \end{equation}
        }
        \footnotesize
        \textbf{\underline{HW Specification:}} \tcb Trigger [T2]
        {\tiny
        \begin{equation}
        \begin{split}
          &\mathit{log_{full} \lor reset \rightarrow flush}
        \end{split}
        \end{equation}
        }
    }
}
\caption{Hardware Spec.: Log Monitor Sub-Module}
\label{fig:seclog}
\end{figure}

$\bullet$ \textbf{Log Monitor: } The Log Monitor tracks the size of \log ($\ptr$) and controls the \textbf{[T2]} \trigger (activated when \log is full). In Figure~\ref{fig:hw}, \textbf{[T2]} is represented by the \textit{flush} hardware signal, which is controlled by Log Monitor and used as an NMI input to the CPU. Thus, when $flush=1$, it immediately launches \tcb execution. Log Monitor is also responsible for indicating that a control flow transfer is ready to be written to \log, by setting the bit $hw_{en}=1$. The $hw_{en}$ bit is an internal wire of the \CFA module used for communication between the Log Monitor and the Logger sub-modules.
Figure~\ref{fig:seclog} details the logic implemented to control $\ptr$ and $\mathit hw_{en}$. It compares $\ptr$ to the maximum size of \log to determine if \log is full, and the result of this comparison determines if $\mathit log_{full}$ bit must be set. Whenever a branch is detected during \textit{\er}  execution ($PC \in \er$) and \log is not full, it sets $\mathit hw_{en}=1$; otherwise, $hw_{en}$ remains 0. 
The same bit is set to $1$ when \er is invoked, which is detected by monitoring the next $PC$ value ($PC_{next}$).
%
In addition, Log Monitor is responsible for incrementing $\ptr$ to keep track of the next unused position in \log.
It ensures to clear $\ptr$ after \tcb execution completes ($PC = TCB_{max}$).
Therefore, upon returning from \tcb, \log will be overwritten by subsequent control flow transfers.
The $loop_{detect}$ signal is used for optimization purposes and controlled by the Loop Monitor sub-module that will be discussed next. 
%
%

\begin{figure}[t]
\footnotesize
\fbox{
    \parbox{0.95\columnwidth}{
        \textbf{\underline{HW Specification:}} Log Entry Construction
        \begin{equation}
        \tiny
        src =
        \left\{
            \begin{array}{cc}
                PC_{prev}, & \text{if } \mathit{\neg loop_{detect}}\\
                ctr[31:16], & \text{if } \mathit{loop_{detect}}
            \end{array}
        \right.
        \end{equation}

        \begin{equation}
        \tiny
        dest =
        \left\{
            \begin{array}{cc}
                PC, & \text{if } \mathit{\neg loop_{detect}}\\
                ctr[15:0], & \text{if } \mathit{loop_{detect}}
            \end{array}
        \right.
        \end{equation}

        \textbf{\underline{HW Specification:}} Secure Log Update
        \begin{equation}
        \tiny
          hw_{en} \rightarrow \log [\ptr] = (src, dest)
        \end{equation}
    }
}
\caption{Hardware Spec.: Logger Sub-Module}
\label{fig:logger}
\end{figure}

$\bullet$ \textbf{Loop Monitor:} The Loop Monitor is used to reduce \log size by optimizing repetitive \log entries produced by static loops without internal branching instructions (such as delay loops, which are common in embedded system software). These loops can quickly fill \log with redundant control flow transfers.  
Thus, \acron follows a similar approach to prior \CFA methods~\cite{lofat,dessouky2018litehax} by considering each repeated backward jump with the same source and destination addresses (\textit{src, dest}) as a static loop and simply logging (\textit{src, dest}) once, along with the number of iterations that occurred. To differentiate between \log entries generated by static loops from regular entries, Loop Monitor controls the $loop_{detect}$ signal. Since this feature is strictly used for optimization purposes, we defer its details to Appendix~\ref{apdx:optimization}. 
All control flow transfers in a static loop ($loop_{detect} = 1$) do not increment $\ptr$ but instead write the number of iterations in place. Once the control flow leaves the static loop, $\ptr$ is incremented again.

$\bullet$ \textbf{Logger:} The sole responsibility of the Logger module is to write the next entry into the \log. Each entry is a pair of source and destination addresses. Since MSP430 uses 16-bit addresses, entries are 32-bit values. 
Figure~\ref{fig:logger} shows the logic to append an entry to \log.
Whenever $\mathit{hw_{en}}=1$, 
Logger appends ($\mathit{src=PC_{prev}, dest=PC}$) to \log, where $PC_{prev}$ denotes the previous $PC$ value. 
If a static loop is detected ($\mathit{loop_{detect}}=1$), the loop counter \textit{ctr} is additionally logged ($\mathit{src=ctr[31:16], dest=ctr[15:0]}$) as a next \log entry and incremented accordingly. 
%

In summary, the CFA Module supports security properties \textbf{[P1-P4]}. The Boundary Monitor supports both \textbf{[P1-P2]} to ensure critical data cannot be modified maliciously. In combination, Log Monitor, Branch Monitor, Loop Monitor, and Logger detect and record all control flow transfers to \log, implementing \textbf{[P3]}. Furthermore, \textbf{[P4]} is supported by the Log Monitor ensuring \textbf{[T2]} will always cause TCB to execute.\\

\vspace{-0.6em}
\begin{center}
 {\bf Active RoT Module}
\end{center}
\vspace{-0.6em}

Properties {\bf[P4-P5]} also rely on the active RoT guarantees discussed in Section~\ref{sec:garota}. The sequence \tcb-Att$\rightarrow$\tcb-Wait$\rightarrow$\tcb-Heal is implemented within that active RoT handler function \func.
To ensure that triggers {\bf[T1]}, {\bf[T2]}, and {\bf[T3]} always result in the execution of \func, they are NMIs (that cannot be disabled in software) thus supporting {\bf[P4]}.
In addition, \acron hardware makes use of some of the original hardware modules in \garota~\cite{garota} implementation to ensure \textbf{\textit{Non-malleability}} and {\bf \textit{Atomicity}} of \tcb, thus supporting {\bf[P5]}.
\vrased is ported in its entirety into \acron. It is required to implement the \textit{\tcb-Att} phase of the \textit{\tcb} sequence securely, i.e., securely producing an \RA measurement and authenticating \vrf messages, per property {\bf [P5]}. \vrased verified architecture enables secure \RA by ensuring confidentiality of the attestation key and proper execution of the \RA integrity-ensuring function.
It assures that untrusted software in \prv cannot access any trace of \attkey before, during, or after \tcb-Att computation.
In addition, \vrased ensures that \textit{\tcb-Att} remains immutable, executes atomically, and has fixed entry and exit points (see~\cite{vrased} for details on \vrased internals).


\subsection{Security Analysis}
\label{sec:security}

Recall from Section~\ref{sec:adv_model} that \adv can exploit vulnerabilities in \prv software $S$, to modify any code or data (including stack data to perform control flow hijacks). Similarly, \adv may use this capability to disrupt any phase of \acron protocol.
%

\adv may attempt to forge/modify \log to reflect the control flow path of \er faithful execution without truly executing it. One approach is to modify \log directly. However, as \acron prevents all CPU/DMA accesses to \log (\textbf{[P1]}), any such attempt results in a system reset, triggering \tcb to inform \vrf of this attack. In addition, \adv may attempt to forge \log by causing \acron to track a different executable \textit{$\er_{\adv}$}, located elsewhere in $PMEM$, by modifying the bounds in $METADATA$. However, \textbf{[P2]} assures that such an attempt to modify $METADATA$ is prevented. Similarly, \adv cannot overwrite \chal in an attempt to replay the \CFA report produced by a previous execution of \er.

If \adv exploits vulnerabilities to diverge \textit{\er} control flow, the exploit will be reflected in \log (given {\bf [P3]}). Then, to escape \vrf detection, \adv must directly forge an authenticated \CFA report containing a benign control flow path. In order to forge this report, \adv must forge the result of the authenticated integrity-ensuring function (MAC) computed by \tcb-Att, which in turn requires tampering with \tcb execution. 
However, this is infeasible due to \textbf{[P5]}.

\adv may try to cause a deadlock in the \CFA protocol so that \vrf-issued remediation is never performed on \prv. Since \textit{\tcb} is automatically triggered on $\er$ completion, \adv may continuously interrupt \er to prevent \textit{\tcb} from ever being called. Additionally, \adv may overwrite data to cause an infinite loop within $\er$ so that its execution never completes. However, this will be preempted by either: \acron timer \trigger or \acron \log full \trigger. Since \textit{\tcb} is guaranteed to execute thereafter (\textbf{[P4]}), \adv cannot avoid detection/remediation in this way.

\adv may attempt to tamper with \tcb behavior by changing its code or interrupting its execution (e.g., to prevent the remediation phase in \textit{\tcb-Heal}). 
However, \tcb code is immutable at runtime and its execution is atomic (\textbf{[P5]}).
Any attempt to interrupt \tcb will cause \prv to reset and re-execute \tcb from the start (due to \textit{re-triggering on failure} property). After a reset, interrupts and DMA are disabled by default. Therefore, no further interference by \adv is possible during the new instance of \tcb execution.

Finally, a network \adv may discard messages between \vrf and \prv. \acron guarantees that \prv remains in the \textit{\tcb-Wait} phase until an approval message from \vrf is eventually received. Therefore, even when \adv controls both \prv and the network, a compromised \er is prevented from executing.

\section{\reqchange{1}{Alternative Designs \& Policies}}\label{sec:alternatives}

\blue{This section discusses alternatives in \acron design and policies, as well as their implications.}

\reqchange{5}{{\bf $\bullet$ \acron with Hardware Hash Engines:} \acron standard design opts for a software implementation of the \RA integrity-ensuring function (using VRASED). This choice significantly reduces the hardware cost. However, it also increases storage requirements in order to maintain \log verbatim. It also increases the potential number of partial transmissions of \log to \vrf, when this dedicated storage fills up. We note that, if \prv is implemented with a less strict hardware budget, hardware-based hash engines can be utilized to reduce the storage/transmission overhead. In practice, trade-offs between hardware cost and other overheads should be considered when deciding for one approach over the other. Early \CFA methods~\cite{cflat,lofat,dessouky2018litehax,zeitouni2017atrium} proposed to build \log as a hash-chain to compress the sequence of control flow transfers into a single hash digest.
In that way, \prv is only required to store the current hash-chain digest and extend it with the next control flow transfer as it occurs. 
While this can be obtained in a relatively easy way (assuming hash engines are available), it also requires \vrf to enumerate all control flow paths (for the entire execution) that could have led to the received final hash digest. Unfortunately, the complexity of this task grows exponentially with the number of control flow transfers in the path, leading to the well-known path explosion problem~\cite{aliasing,baldoni2018survey,oat,tinycfa}. With these trade-offs in mind, we also provide an implementation of \acron using a hardware hash engine and compare its overhead with the original \acron design. This implementation and comparative results are discussed in Section~\ref{sec:impl}.}

\reqchange{2}{{\bf $\bullet$ Strict vs. Best-Effort Auditing and Remediation:} as discussed in Section~\ref{sec:design_overview}, we describe the strict version of \acron protocol, in which auditing software integrity is a first-class priority (e.g., consider an MCU deployed as a part of a nuclear facility). Therefore, whenever a \CFA report is sent to \vrf, \tcb in \prv waits for a response indefinitely (while re-transmitting the report periodically). In that case, to avoid detection, a Dolev-Yao \adv might jam the network communication rendering \prv unavailable. We note that resuming the execution of the attested application in this case is entirely possible (i.e., by jumping from step 2 to step 9 in Figure~\ref{fig:prot} upon a timeout). While this may be desired in some application domains, it remains unclear why one would aim to guarantee availability to a compromised application. Alternatively, \tcb could resume \er execution for a fixed finite period, issuing a subsequent timer \trigger to check if \vrf response was received. In the latter, \prv does not ``busy-wait'' on \vrf response. The security implications of these policies to the application domain should be considered carefully.}

\reqchange{4}{{\bf $\bullet$ Adapting \acron to Higher-End Devices:} as noted in Section~\ref{sec:scope}, \acron initial design targets bare-metal MCUs. Adapting \acron for higher-end systems is an interesting and promising direction for future work. The main challenge lies in the dependence of higher-end systems on MMU-based virtual memory assignment, whereas \acron performs its checks based on physical addresses. On higher-end devices, the MMU translations are themselves controlled by privileged software that could be itself compromised and tamper with address translations to circumvent \acron. Future work could consider methods to verify the consistency of virtual-to-physical address translations across the runtime of an attested/audited process, perhaps by augmenting MMUs with new (yet backward-compatible) hardware features.}

{\bf $\bullet$ Non-Control Data Attestation:} Subtle software integrity attacks are still possible by compromising non-control data, without modifying a program's control flow path. While this class of vulnerabilities (e.g., ``write anywhere'' vulnerability) is less common, they are still possible. One approach to deal with this problem is to append all data inputs (any memory read from outside the attested program's stack) to \log, as proposed in~\cite{dialed}. In possession of both the executed control flow path and all data inputs, \vrf can abstractly execute the attested program~\cite{oat} to observe any such exploit.

\section{Implementation \& Evaluation}
\label{sec:impl}

\acron implements the workflow and architecture shown in Figures~\ref{fig:phases} and~\ref{fig:system}, respectively. As discussed in Section~\ref{sec:scope}, our prototype is built on the low-end MSP430 MCU, primarily due to its simplicity and open-source availability.

We use Xilinx Vivado tool-set to synthesize \acron hardware. \acron hardware is written in the Verilog hardware description language and implements each of \acron sub-modules according to the logic defined in Section~\ref{sec:design}.
In total, \acron hardware is implemented in 2042 lines of Verilog code. The \CFA module accounts for 982 lines, the Active RoT module (including VRASED) accounts for additional 927 lines, and 123 lines tie the two modules together. 
We then synthesized and deployed \acron on the Basys3 prototyping board that features an Artix-7 FPGA. 

The \tcb Software implements functions, \tcb-Att, \tcb-Wait, and \tcb-Heal, as described in Section~\ref{sec:design}. All three functions are linked so that the entire \tcb code is located within a contiguous region of $PMEM$. This ensures that \tcb can be monitored and protected as intended by the \acron hardware.
%

\tcb-Wait is responsible for communicating with \vrf and authenticating \vrf messages. We use \vrased authentication module to support \vrf authentication in \tcb-Wait. 
In our prototype, \prv and \vrf are physically connected using a USB-UART interface.
As explained in Section~\ref{sec:design_overview}, \tcb-Heal is configurable to meet application needs. 
In our prototype and evaluation, we implement a simple remediation option: shutting down \prv.

\reqchange{6}{
To assess the trade-off discussed in Section~\ref{sec:alternatives}, we also implement an \acron variant using a hash engine. It integrates a SPONGENT hash engine implemented for openMSP430 by the SANCUS project~\cite{Sancus17}.
In this variant, the hash engine module replaces the Logger module and receives the same inputs, i.e., ($src, dest$) and $hw_{en}$. 
When $hw_{en}$ is set, the ($src, dest$) pair is accumulated into the SPOGENT hash digest. 
The hash engine operates with default parameters: at 100MHz in a 128-128-8 bit SPONGENT configuration, producing a 128-bit digest. 
Since each control flow contains 32 bits of data (16-bit source and destination addresses), the hash engine reads 8 bits at a time from a 512-bit FIFO buffer of control flow transfers at each cycle. With the hash engine, the Logger module is no longer required. Similarly, as \vrf is not provided with \log verbatim (see Section~\ref{sec:alternatives} for a discussion on implications related to path explosion), the $flush$ signal and the tracking of \log size are not required. The hash engine (including its integration with other \acron modules) is implemented in 730 lines of Verilog code.
}

\subsection{Results}

To the best of our knowledge, no prior work implements \acron features. Nonetheless, to provide a reference point, we report \acron costs in comparison to other security architectures targeting the same class of MCUs (namely VRASED~\cite{vrased}, SANCUS~\cite{Sancus17}, GAROTA~\cite{garota}, and Tiny-CFA~\cite{tinycfa}).
We also compare \acron to closely related hardware-based \CFA architectures (namely LiteHAX~\cite{dessouky2018litehax}, LoFAT~\cite{lofat} and Atrium~\cite{zeitouni2017atrium}) noting that these architectures were implemented on a different MCU class with a less strict hardware budget than the MSP430.

\vspace{-0.6em}
\begin{center}
 {\bf Hardware and Memory Overhead}
\end{center}
\vspace{-0.6em}

\begin{figure}[t]
    \centering
    \subfigure[MSP430-based implementations]{\label{fig:hw_msp430}\includegraphics[width=0.9\columnwidth]{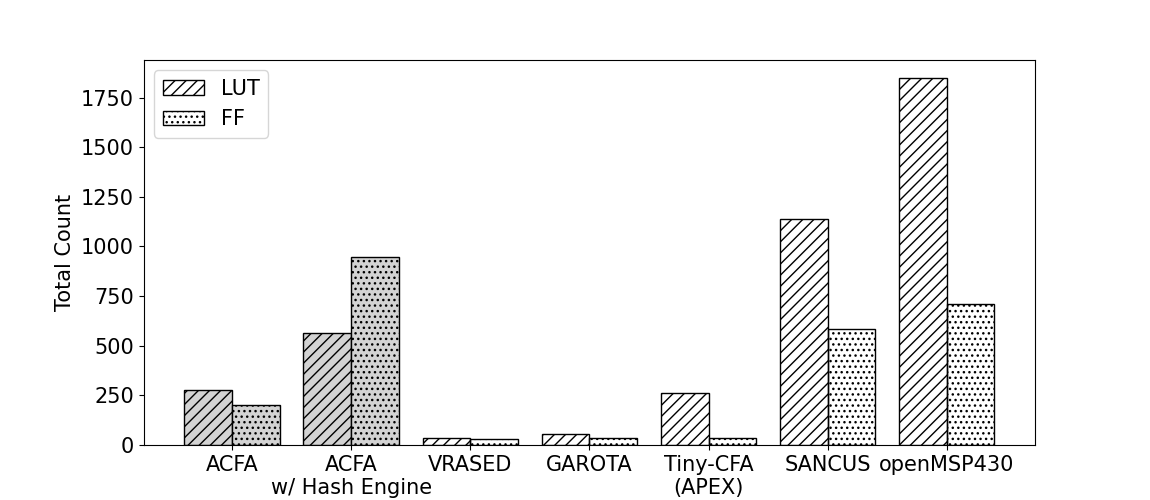}}
    \subfigure[\CFA-specific architectures]{\label{fig:hw_hwb}\includegraphics[width=0.9\columnwidth]{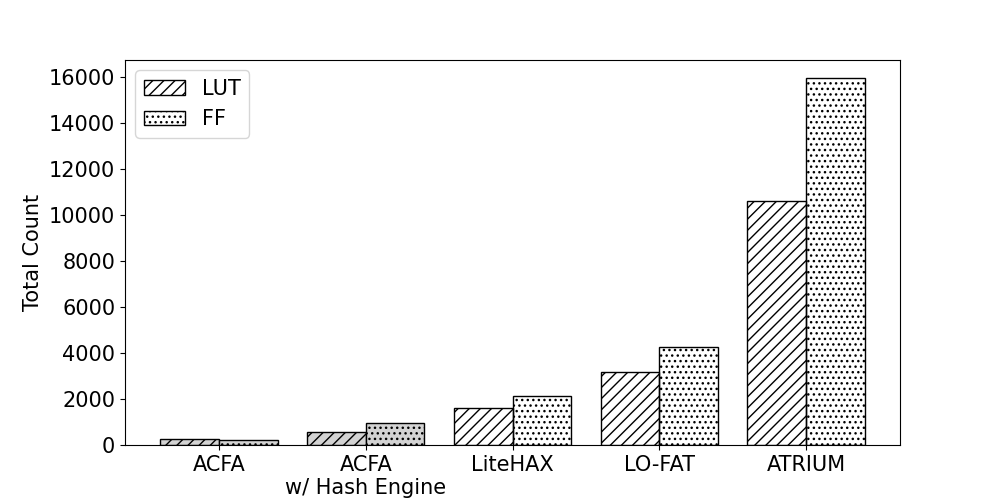}}
     \vspace{-1em}
    \caption{\reqchange{6}{Comparison of hardware cost}}
    \label{fig:hwcost}
\end{figure}

Similar to the related work, we consider the hardware overhead in terms of additional Look-up Tables (LUTs) and flip-flops/registers (FFs). The increase in LUTs estimates the additional chip cost and size due to combinatorial logic. The number of extra FFs indicates additional state required by sequential logic. Figure~\ref{fig:hwcost} compares \acron hardware cost with other architectures and a baseline unmodified openMSP430 core. Overall, \acron requires additional 275 LUTs and 202 FFs. This represents a $\approx$18.7\% increase with respect to the openMSP430 core.

Due to its hybrid design, the standard version of \acron incurs significantly lower hardware overhead than hardware-based approaches such as SANCUS, as shown in Figure~\ref{fig:hw_msp430}.
Compared to other hybrid architectures, i.e., Tiny-CFA/APEX, \acron incurs the similar number of LUTs but requires more FFs for sequential logic. Since \acron is a superset of \vrased and \garota, \acron hardware cost is naturally larger than the two.

Figure~\ref{fig:hw_hwb} compares \acron with hardware-based \CFA architectures, showing that a hybrid design for \CFA requires less hardware.
For instance, \acron requires $\approx$5.8 times less LUTs and $\approx$10.5 times less FFs than LiteHAX, which is the cheapest related hardware-based \CFA architecture. 

\reqchange{6}{Finally, the \acron variant equipped with a hash engine adds 510 LUTs and 946 FFs to the openMSP430 baseline, representing an increase of 235 LUTs and 744 FFs over \acron standard design. 
This difference highlights the hardware savings of a hybrid \CFA approach. 
Nonetheless, it is important to note that a hash engine reduces the storage/transmission overhead of \log on \prv. On the other hand, it also increases verification complexity (to be performed by \vrf) exponentially due to the path explosion problem~\cite{oat,tinycfa,cflat,scarr}. These trade-offs should be considered carefully when deciding for a particular design option.}

\vspace{-0.6em}
\begin{center}
 {\bf Runtime Overhead}
\end{center}
\vspace{-0.6em}

Since \acron does not require code instrumentation, no runtime overhead is incurred to save entries to \log. Similarly, there is no code size increase for \er. 
The exact runtime overhead incurred by the \tcb execution varies depending on factors such as communication delays, \vrf choice of remediation function, and time taken by \vrf to verify reports.
In practice, however, when testing our end-to-end application use-cases, we have noticed that this overhead is 
dominated by the time required to compute the HMAC function on \prv and to communicate between \prv and \vrf. 
Because of this, \vrf should consider a suitable configuration of \log and $AER$ sizes depending on their response time requirements.
We discuss more details and timing results for the end-to-end prototype in Section~\ref{sec:end_to_end}.

\vspace{-0.6em}
\begin{center}
 {\bf Evaluation with Sample Applications}
\end{center}
\vspace{-0.6em}

We evaluate \acron on three exemplary applications (which were ported to run on openMSP430): an Ultrasonic Sensor~\cite{ultsensor}, a Temperature Sensor~\cite{tempsensor}, and a Syringe Pump~\cite{opensyringe}.

During evaluation of the application software, we fix the timeout period (for \trigger \textbf{[T1]}) to 50ms. We note that in practice we expect this time-out to be much larger. However, we choose a small value to force \trigger occurrences so as to evaluate a worst-case.  We consider two maximum \log sizes: 0.5kB and 1.0kB (similar to the timer we intentionally choose very small \log sizes to force \trigger-s).
The boundaries of \er are fixed to cover the entire untrusted software: $\er = S$. 
Table~\ref{tab:trigger_stats}
shows the size of \log data (i.e., total \log bytes), the number of \acron reports sent to \vrf, and the number of \trigger-s issued during execution of each sample application under both maximum $\ptr$ settings. 

%

\begin{table}[t]
\centering
\caption{Runtime statistics for 0.5kB and 1 KB \log}
\label{tab:trigger_stats}
\resizebox{\columnwidth}{!}{%
\smaller
\begin{tabular}{|c|c|c|c|c|c|c|}
\hline
\textbf{Program} & \textbf{Max $\ptr$} & \textbf{\# {[T1]}} & \textbf{\# {[T2]}} & \textbf{\# {[T3]}} & \textbf{\log Data} & \textbf{\# Reports} \\ \hline
\multirow{2}{*}{Ultrasonic Sensor} & 0.5kB & 0 & 0 & 2 & 0.01 KB & 2 \\ \cline{2-7} 
 & 1.0kB & 0 & 0 & 2 & 0.01 KB & 2 \\ \hline
\multirow{2}{*}{Temperature Sensor} & 0.5kB & 0 & 2 & 2 & 1.2 KB & 4 \\ \cline{2-7} 
 & 1.0kB & 0 & 1 & 2 & 1.2 KB & 2 \\ \hline
\multirow{2}{*}{Syringe Pump} & 0.5kB & 0 & 7 & 2 & 3.6 KB & 9 \\ \cline{2-7} 
 & 1.0kB & 5 & 0 & 2 & 3.6 KB & 7 \\ \hline
\end{tabular}
}
\end{table}

\begin{table}[t]
    \centering
    \caption{\acron vs. prior \CFA for MCUs qualitatively}
    \label{tab:comp}
    \resizebox{\columnwidth}{!}{%
    \begin{tabular}{|c|c|c|c|c|c|c|c|c|}
        \hline
         & C-FLAT~\cite{cflat} & LO-FAT~\cite{lofat},LiteHAX~\cite{dessouky2018litehax} & Tiny-CFA~\cite{tinycfa},DIALED~\cite{dialed},OAT~\cite{oat} & \textbf{\acron} \\
        \hline
        SW Instrumentation & Yes  & No & Yes & \textbf{No} \\
        HW Hash Engine & No  & Yes & No & \textbf{No} \\
        \log slicing & No  & No & No & \textbf{Yes} \\
        Control Flow Auditing & No  & No & No & \textbf{Yes} \\
        Active Remediation & No  & No & No & \textbf{Yes} \\
        \hline
    \end{tabular}
    }
\end{table}

The Ultrasonic Sensor application contains very few control flow transfers, so its execution does not fill up \log. Thus, executing this application causes only two \textbf{[T3]} \trigger-s (one at boot and one at the end of execution). 
On the other hand, both Temperature Sensor and Syringe Pump applications produce more control flow transfers, filling up the 0.5kB in \log before their execution is completed. 
Hence, additional \trigger-s occur during their execution to send partial \log reports to \vrf.
For both of these applications, increasing the maximum $\ptr$ results in fewer \trigger-s, since \textbf{[T2]} \trigger will happen less often.
We observe that, for the Syringe Pump application, fewer reports are generated and also fewer \trigger-s occur with a larger \log size. 
In addition, the source of intermediate reports changes from \textbf{[T2]} to \textbf{[T1]}.
This is because all intermediate reports surpass 0.5kB before reaching the 50ms threshold. However, this does not occur when \log size is 1KB.
This small change also causes fewer reports to be generated due to the runtime of the application.

Given these observations, device operators should consider the trade-off between resource allocation (e.g., for storing \log), timeout periods, and application requirements.

\vspace{-0.6em}
\begin{center}
 {\bf Energy Consumption}
\end{center}
\vspace{-0.6em}

Using the Vivado tool-set, we synthesize the hardware and generate energy consumption reports for \acron.
The unmodified openMSP430 requires $0.06$W, whereas \acron hardware requires an additional $0.001$W. This represents a 1.6\% increase in energy consumption.

\vspace{-0.6em}
\begin{center}
 {\bf \CFA Verification}\label{sec:cfa-ver}
\end{center}
\vspace{-0.6em}

Our discussion thus far emphasizes \acron architecture on \prv. Based on the authenticated information produced by this architecture, \vrf must determine if a runtime attack has occurred. This verification can be implemented in a number of ways.
In any case, \vrf must always check the \RA measurement to confirm that the expected binary has not been modified. Similarly, it must examine the size and contents of \log to determine if any violations occurred during the execution and generation of \log.
In our implementation, \vrf generates the CFG of \prv binary to check if the control flow transfers reported in \log match a valid path in the CFG. \vrf also uses a shadow stack to confirm the validity of return addresses reported in \log.
Aside from this sample implementation, any CFI policy that would otherwise be implemented on the resource-constrained \prv can now be outsourced to the higher-end \vrf for faster and less intrusive runtime integrity verification.

\subsection{\reqchange{1,RC4}{End-to-End Prototype}}
\label{sec:end_to_end}

To demonstrate \acron's practicality in on-demand sensing settings,
we implement a fully-functional prototype including \prv and \vrf realizing the \acron end-to-end workflow in real-time. The implementation and a video demonstration of this end-to-end example are also available at~\cite{acfarepo}.

{\bf \prv setup:}
In this application, \prv contains a simple program that receives a password input from a remote user and compares it with an expected password.
We intentionally introduce a buffer-overflow vulnerability in this program by not performing array bound checks when storing the received password input.
As a result, \adv can overflow the buffer and overwrite a return address.
If the correct password is entered, \prv then records six ultrasonic-sensor readings and exits the program.
In terms of \acron configuration parameters, we set the maximum \log size to 256B and the timeout period to an overwhelmingly large value, essentially deactivating timeout triggers (\textbf{[T1]}).

\blue{
{\bf \vrf Offline Phase (performed once):}
We implement \vrf in Python and execute it on a 64-bit Ubuntu 18.04 machine with an Intel i7 @ 3.6GHz. 
\vrf offline phase consists of two main tasks.
First, \vrf pre-computes the \prv program's CFG 
by parsing the object-dump file of the application binary.
In addition, \vrf prepares to verify the \acron report (Step 4 in Figure~\ref{fig:prot}) by pre-computing the hash of $AER$.
Since $AER$ is not expected to change between the reports, this optimization is put in place to speed up the online verification process.
The entire offline phase takes $\approx$2.5s.
}

\blue{
{\bf \vrf Online Phase (performed on every protocol instance):}
In this phase, \vrf receives \log slice from \prv and aims to validate whether \log corresponds to a valid software execution, i.e., following a specific path in CFG generated from the offline phase.
Each received \log slice may be acquired from the first, intermediate and last \acron reports. \vrf validates \log from each type of report as follows:}
\blue{
\begin{compactitem}
    \item \textbf{First \acron report.} Recall that a first \acron report is valid if it is produced on \prv immediately after \er is invoked. Thus, in this phase, \vrf checks whether its first entry corresponds to a function call to \er, i.e., the destination address matches $\er_{min}$. The rest of \log entries must adhere to a valid control flow path in the CFG generated in the offline phase.
    \item \textbf{Intermediate \acron reports.} 
    In all reports other than the first and the last, the first \log entry must correspond to a jump from $\tcb_{max}$ to \er after obtaining \vrf approval for the previous report. Subsequent entries must continue a valid control flow path. 
    Hence, \vrf checks the first entry by comparing its source address to $\tcb_{max}$ and its destination to \er region.
    To validate other entries, \vrf keeps traversing CFG.
    \item \textbf{Last \acron report.}  \er execution completion results in a \trigger to the \tcb. The first \log entry in this report is checked in the same way as in an intermediate report. \vrf validates the last \log entry by matching it with a control flow transfer caused by a jump from $\er_{max}$ to $\tcb_{min}$. For other \log entries, \vrf continues to traverse the CFG.
    \item \textbf{Single \acron report.} In the case that no intermediate reports are created, \vrf obtains a single \acron report that covers the entire \er execution trace. \vrf checks validity of this trace via CFG traversal.
\end{compactitem}
}

\blue{
In addition, to detect attacks overwriting return addresses, 
we implement a shadow stack on \vrf during the online phase.
While traversing the CFG, if \vrf encounters a function call, it pushes the expected return address to the shadow stack. 
Upon returning from a function, \vrf pops the return address from the stack and matches it with the corresponding \log entry obtained from \prv. 
}

\blue{
{\bf Remediation Options:}
%
Some remediation options can be implemented with very few lines of code, such as a system shutdown. 
In MSP430, for instance, this can be implemented by simply setting a bit in the system status register. 
Similarly, system reset can be achieved by calling the reset vector (the first address of the interrupt vector table). 
In other cases, \vrf may implement a more thorough remediation option, such as erasing data memory or updating the MCU software followed by a system reset. 
Our end-to-end example supports both shut-down and reset options. 
In our sample implementation, \vrf detects an invalid \log in the second instance of the \acron protocol and thus successfully discovers the buffer-overflow attack on \prv. 
In this case, \vrf chooses to remediate \prv by shutting it down, preventing the malicious program from running on \prv.
}

\blue{
{\bf Timing Results: } We evaluate the end-to-end sample by timing the protocol for a varied size of $AER$. When the end-to-end example runs with the benign input, four 256-Byte partial \log-s are generated and transmitted to \vrf. Table~\ref{tab:demo_timing} presents the average runtime of each step in the protocol of Figure~\ref{fig:prot}. The overall runtime of the protocol increases as the size of the $AER$ increases since more time is required to attest a larger region of program memory. The remaining steps are completed in constant time as they do not depend on the size of $AER$ -- including the time to verify \acron report due to the pre-computation of \er hash in \vrf offline phase. Steps 1-3 and Step 6 require the most time due to the communication delay.
}

\begin{table}[]
\caption{End-to-end protocol timing (in ms) for varying size of $AER$. Steps correspond to the protocol steps of Figure~\ref{fig:prot}}
\label{tab:demo_timing}
\resizebox{\columnwidth}{!}{
\begin{tabular}{|c|c|c|c|c|c|c|}
\hline
\textbf{\textit{AER} Size} & \textbf{Steps 1-3} & \textbf{Step 4} & \textbf{Step 5} & \textbf{Step 6} & \textbf{Steps 7-8} & \textbf{Total} \\ \hline
1 KB      & 351.9     & 2.096  & 6.70E-02 & 205.8  & 96.05    & 655.9 \\
2 KB      & 367.9     & 1.887  & 7.12E-02 & 206    & 96.01    & 671.9 \\
4 KB      & 399.8     & 2.557  & 8.24E-02 & 205.2  & 96.1     & 703.8 \\
8 KB      & 463.8     & 2.054  & 7.66E-02 & 205.8  & 96.12    & 767.9 \\ \hline
\end{tabular}
}
\end{table}


Table~\ref{tab:demo_timing} also shows the runtime of cryptographic computations used to authenticate messages. 
In Step 5, \vrf produces $Auth$ with an average runtime of $\approx0.07$ms. In Step 7, \prv produces $out$ with an average runtime of $\approx96.1$ms. Step 8 (either a call to \tcb-Heal or return to \textit{S}) requires negligible time. Thus, \prv time to produce a MAC dominates the reported runtime of Steps 7-8.

This end-to-end example aims to illustrate the impact of more complex software on \acron's workflow. Including the receipt of network inputs, the password-check, and 6 sequential sensing operations, this sample application incurs $\approx 6,000$ control flow transfers. While \acron guarantees are maintained as the complexity of applications increases, the number of communication rounds in \acron pipeline (and associated overhead) also increases accordingly.



\vspace{-1em}
\section{Related Work}
\label{sec:relatedwork}
\vspace{-0.6em}

\textbf{RA:} \RA architectures fall into three categories: software-based, hardware-based, or hybrid. Software-based \RA~\cite{KeJa03, SPD+04, SLS+05, SLP08} does not depend on specialized hardware modules and does not require any modifications to the existing hardware on a device. However, these approaches are limited due to their reliance on strong assumptions about adversaries' capabilities and timing requirements for the link connecting \vrf and \prv. Hardware-based methods~\cite{PFM+04, KKW+12, SWP08, DAA} use dedicated hardware support either from external modules such as TPMs~\cite{tpm} or from the instruction set architecture, as in Intel SGX~\cite{sgx,sgx-explained}.
\acron leverages hybrid \RA architecture VRASED~\cite{vrased} to implement a part of its active RoT for \CFA, responsible for measuring the installed binary and authenticating \log before a report can be sent to \vrf.
In addition to VRASED~\cite{vrased}, other hybrid \RA approaches such as SMART~\cite{smart} and TyTAN~\cite{tytan} use a combination of software and hardware for attestation. Typically, the hardware cost of hybrid approaches is substantially lower because they implement the \RA  measurement (e.g., MAC or signature) in software, while a hardware monitor is used to validate \prv execution, ensuring the integrity of the \RA execution and the secrecy of the \RA cryptographic key(s).
RealSWATT~\cite{surminski2021realswatt} is a recent software-based approach to  continuously attest real-time and multi-core systems. It dedicates a core to attesting other applications continuously. In contrast, ACFA (and other hybrid architectures) targets single-core bare-metal MCUs, where RealSWATT would not apply. Different from the above-mentioned static \RA methods, \acron aims to support secure control flow \textit{auditing}, in addition to attestation.

\textbf{CFI Methods:} CFI~\cite{stackguard,Abadi2009} is a class of approaches (we include shadow stacks~\cite{burow2019sok} and Address Space Layout Randomization (ASLR)~\cite{aslr} in this class) intimately related to \CFA. While CFI has the similar goal of ensuring that a valid program path has been executed, \CFA is more suitable for resource constrained devices. Compared to \CFA, \textit{CFI} does not provide \vrf with \log and instead checks the control flow locally -- on \prv. In addition, many \textit{CFI} methods rely on security capabilities that are usually expensive to low-end MCUs (e.g., MMUs). \CFA, on the other hand, outsources the control flow verification to \vrf: a more resourceful trusted device. 
GRIFFIN~\cite{ge2017griffin} uses a shadow stack to restrict return targets. It also restricts indirect call sites and does not log static transfers, reducing storage requirements. Similar optimizations could be applied to \acron to reduce \log size.
In general, CFI is considered challenging due to the hardness of associated sub-problems. For a discussion on CFI, see~\cite{war_in_mem}.

\textbf{Runtime Attestation \& CFA Methods:} C-FLAT~\cite{cflat} was the earliest work on \CFA. It relies on binary instrumentation along with hardware support from ARM TrustZone~\cite{trustzone} to securely log control flow transfers in TrustZone's protected memory. At each instruction that alters the control flow (e.g., jump, branch, return), execution is trapped into the secure world and the control flow path taken is logged into protected memory.
LO-FAT~\cite{lofat} and LiteHAX~\cite{dessouky2018litehax} are custom hardware-based approaches that improve upon C-FLAT by removing the need for binary instrumentation and by moving away from TrustZone. They introduce custom hardware support to hash branching instructions at runtime. As a result, instrumentation is no longer required and the runtime overhead (both execution time and code size) is reduced. However, an expensive hardware overhead is incurred due to the introduction of a hardware hash engine. Similar to C-FLAT, Tiny-CFA~\cite{tinycfa} also relies on instrumentation, but leverages cheaper hardware support from the Proof-of-Execution architecture APEX~\cite{apex}. Therefore, it provides \CFA at a relatively lower cost, making \CFA amenable to low-end MCUs. Unlike prior architectures, Tiny-CFA constructs a verbatim log of control flow transitions, rather than computing a hash-chain. Therefore, Tiny-CFA is limited by the growth of \log in relation to the amount of memory available to store it on \prv. Nonetheless, this approach benefits from not requiring \vrf to enumerate all possible valid control flow paths. DIALED~\cite{dialed} builds upon Tiny-CFA to also provide Data Flow Attestation (DFA). Similarly, OAT~\cite{oat} augments a variant of C-FLAT with DFA and provides optimizations to reduce the size of \log, when sent to \vrf verbatim. Compared to \CFA, DFA~\cite{oat,dessouky2018litehax} also detects ``non-control data-only attacks'' that corrupt intermediate data memory values during execution without affecting the program's control flow. While vulnerabilities that enable this type of attack are less common, they are still possible in specific cases (see~\cite{dialed} for examples).

\textbf{Comparison of \acron with Related Work:} \acron addresses key limitations of prior \CFA methods.
To the best of our knowledge, \acron is the first \CFA technique to support secure control flow auditing and remote remediation guarantees when control flow attacks are detected by \vrf.
It also supports streamed reports that slice \log, making continuous \CFA possible to large or infinite executions.
In addition, \acron implements the first hybrid approach to simultaneously obviate the need for instrumentation and minimize \CFA hardware overhead.
Table~\ref{tab:comp} presents a qualitative comparison between \acron and prior \CFA architectures. \acron does not incur overhead due to instrumentation or hardware hash engines. It also constructs fixed size reports that are continuously streamed to \vrf. Finally, unlike prior \CFA, \acron supports active remediation when \vrf determines that \prv has been compromised, as well as control flow auditing capabilities.

\ignore{
\edit{
\textbf{\vrased and \garota in \acron.}
Hardware components from GAROTA and VRASED support the \acron's ability to have active remediation and secure attestation.
Without modification and additional hardware, \vrased and \garota do not achieve the requirements of secure control flow auditing with guaranteed remediation. \acron addresses several architectural challenges that, to our knowledge, haven’t been attempted -- including in \vrased and \garota.
We believe ACFA is the first architecture to propose secure interrupts to detect and handle control flow events on MCUs. These control flow-based interrupts require novel hardware modules unique to \acron to detect and log events of interest. On the other hand, \vrased and \garota are completely oblivious to control flow events. \acron requires composition of VRASED and GAROTA while \textit{not requiring all of their components}.
The \acron remediation function to provide control flow guarantees is unique, while making use of active-RoT concept proposed in the GAROTA. \acron makes use of GAROTA similarly to many system papers that leverage existing architectural support (e.g., SGX or TrustZone) to provide novel security guarantees.
}}
\section{Conclusion}
\label{sec:concl}

We designed, implemented, and evaluated \acron: an inexpensive hybrid active \CFA architecture that supports control flow auditing and guaranteed remediation of detected compromises.
\acron implementation is systematically de-constructed into sub-modules that jointly enforce \acron required properties.
Based on this set of properties, we argue \acron's security.
\acron public prototype (available at ~\cite{acfarepo}) was implemented and synthesized on top of the low-end openMSP430 MCU.

\section*{Acknowledgments}
We sincerely thank the paper's anonymous shepherd and the anonymous reviewers for their constructive comments and feedback.
The first and third authors were supported by the National Science Foundation (Award \#2245531) as well as a Meta Research Award (2022 Towards Trustworthy Products in AR, VR, and Smart Devices RFP).
The second author was supported by the ASEAN IVO (\url{www.nict.go.jp/en/asean_ivo/}) project, Artificial Intelligence Powered Comprehensive Cyber-Security for Smart Healthcare Systems (AIPOSH), funded by NICT (\url{www.nict.go.jp/en/}).

\begin{small}
\bibliographystyle{plain}
\bibliography{references}
\end{small}

\appendix
\section*{APPENDIX}

\section{Detecting Interrupts Accurately}
\label{ap:callirq}

It is possible that several instructions execute in the time between an interrupt being triggered and the CPU actually jumping to the associated ISR. Therefore, Branch Monitor tracks the \textit{irq} signal to determine when an interrupt is triggered and the \textit{gie} signal to determine the moment it is accepted. The signal $\mathit{call_{irq}}$ is set when this pattern is detected. 

In \acron, the signal $\mathit{call_{irq}}$ is an internal signal to Branch Monitor that is set by monitoring \textit{irq} and \textit{gie}. The $\mathit{call_{irq}}$ signal is controlled by the FSM shown in Figure~\ref{fig:irqbranch} within Branch Monitor. When an interrupt is triggered, several cycles take place in order for the MCU to retrieve the address of the interrupt service routine and accept the interrupt. During the time between the interrupt being triggered and actually accepted, it is possible that multiple instructions are executed by the CPU. Therefore this FSM within Branch Monitor is crucial in order to determine the exact instruction that is the source of the transition.

\begin{figure}[th]
\begin{center}
\noindent\resizebox{0.6\columnwidth}{!}{%

    \begin{tikzpicture}[->,>=stealth',auto,node distance=4.0cm,semithick]
        \tikzstyle{every state}=[minimum size=1.5cm]
        \tikzstyle{every node}=[font=\small]

        \node[state] (0,0) (A) {\shortstack{
                \textit{Wait}\\
                \scriptsize $call_{irq}=0$
                }};
        \node[state] [right=3cm of A] (B){\shortstack{
                \textit{Pend}
                }};
        \node[state] (C) [above left=0.75cm and 1.3cm of B] {\shortstack{
                \textit{Acc}\\
                \scriptsize $call_{irq}=1$
                }};

        \draw[->] (A) edge[below] node{\scriptsize \shortstack{$(irq \land gie) \lor nmi$}} (B);
        \draw[->] (B) edge[above, bend right, right=0.3] node{\scriptsize \shortstack{$irq_{acc}$}} (C);
        \draw[->] (C) edge[above, bend right] node{} (A);
    \end{tikzpicture}
}
\caption{Branch Monitor FSM to detect branching due to an interrupt and support \textbf{[P3]} in \acron.}
\label{fig:irqbranch}
\end{center}
\end{figure}

Branch detection due to an interrupt is modeled as a three-state FSM with states $Wait$, $Pend$, and $Acc$ with $Wait$ being the initial state. A transition from $Wait$ to $Pend$ represents the moment a maskable interrupt ($irq$) or 
\acron-specific non-maskable interrupt ($nmi$) due to \textbf{[T1-T3]} has been triggered.  Then, a transition from $Pend$ to $Acc$ occurs when the interrupt is accepted, which is indicated by an internal signal $irq_{acc}$.
After transitioning to $Acc$, $call_{irq}$ is set since a call due to an interrupt has occurred. Once this has been set, the third transition occurs from $Acc$ to $Wait$ and the flag is cleared. Since $call_{irq}$ causes $branch_{detect}$ to be set at this moment (per Figure~\ref{fig:branchdetect}), this allows the log entry for this interrupt to represent the exact instruction that the jump due to the interrupt occurred.

\section{Loop Detection \& Optimization Module}\label{apdx:optimization}

Figure~\ref{fig:loopmon} shows the hardware specification for accurately detecting a loop without internal branches (e.g., delay loops) and counting its iterations. 
Loop Monitor detects a loop based on the current \textit{PC} and the previous PC ($\mathit{PC_{prev}}$). 
It also takes the output signal from Branch Monitor ($\mathit{hw_{en}}$) as input, which determines if a branch has been detected. 

\ignore{
Given a branch is detected, this could potentially be due to a loop. 
However, it is not guaranteed since this could occur due to other instructions, such as a call.
Therefore, the ($src$, $dest$) is saved and an internal counter \textit{ctr} is incremented when this occurs.
}

\begin{figure}
\footnotesize
\fbox{
    \parbox{0.95\columnwidth}{
        \textbf{\underline{HW Specification:}} Loop detection and counting
        \tiny
        \begin{equation}
          hw_{en} \land (ctr==1) \rightarrow (src_{loop} = PC_{prev}) \land (dest_{loop} = PC)
        \end{equation}
        \begin{equation}
          hw_{en} \land (src_{loop} = PC_{prev}) \land (dest_{loop} = PC) \rightarrow (ctr ++)
        \end{equation}
        \begin{equation}
          hw_{en} \land (src_{loop} \neq PC_{prev} \lor dest_{loop} \neq PC) \rightarrow (ctr = 1)
        \end{equation}
        \begin{equation}
            loop_{detect} =
            \left\{
            \begin{array}{cc}
                1, & \text{if } \mathit{ctr > 1}\\
                0, & \text{otherwise}
            \end{array}
            \right.
        \end{equation}
    }
}
\caption{Hardware Spec.: Loop Monitor Sub-module}
\label{fig:loopmon}
\end{figure}

Whenever detecting a branch or $hw_{en}=1$, Loop Monitor saves its source address to $src_{loop}$ signal and its destination address to $dest_{loop}$.
Loop Monitor then uses these signals to detect repeated jumps due to executing a loop. 
When repeated jumps happen $(PC_{prev}, PC) = (src_{loop}, dest_{loop})$,
it increments an internal counter $ctr$ to indicate the number of loop iterations that have occurred.
When $ctr>1$, the Loop Monitor sets $\mathit{loop_{detect}}$ to $1$. 
When the loop execution is over or $(PC_{prev}, PC) \neq (src_{loop}, dest_{loop})$, Loop Monitor resets $ctr$.

The Loop Monitor ensures that all instances of loops are detected and their iterations are counted accurately. Thus, loops are logged to \log efficiently and correctly.


\end{document}

%% file: protocol.tex
\begin{figure*}[t]
	\abovedisplayskip=0pt
	\abovedisplayshortskip=0pt
	\belowdisplayskip=0pt
	\belowdisplayshortskip=0pt
\centering
\scriptsize
\begin{tabular}{|p{.34\textwidth} p{.20\textwidth} p{.36\textwidth}|} 
\hline
\multicolumn{1}{|c}{\textbf{Verifier} (\vrf)} & & \multicolumn{1}{c|}{\textbf{Prover} (\prv)} \\
\hline

& & 1) When \tcb is invoked (either by \trigger \textbf{[T1]-[T3]} or by a manual call in software), \prv executes \tcb-Att to compute \RA measurement:

\begin{equation*}
	H := \cfattest_\attkey(PMEM, METADATA, \log)
\end{equation*} \\

& & where \attkey is the \RA key pre-shared between \vrf and \vrased~\RA RoT in \prv. Then enter \tcb-Wait.\\

3) Receive ($H$, $METADATA$ and \log) and extract $\chal$ from $METADATA$
& \sendmessageleft{top={$\acron$ report}} & 2) In \tcb-Wait: Create and send \acron report $:= H||METADATA||\log$ and wait for approval.\\

& & \\

4) Run verification (including analysis of \log) to determine whether to approve the report:

\begin{equation*}
app:=\vrfy(H, \attkey, PMEM', METADATA, \log)
\end{equation*} & & \\

where $PMEM'$ is the expected software for \prv $PMEM$ and $app \in \{0,1\}$ is an approval bit. & & \\
& & \\
5) Generate a new challenge $\chal'$, a memory region to be monitored ($\er_{min}, \er_{max}$) and an authentication token \auth, where: 

\begin{equation*}
	\auth := \cfattest_\attkey(\chal', \er_{min}, \er_{max}, app)
\end{equation*}

\begin{equation*}
	\chal' := \chal + 1
\end{equation*}
& & \\

6) Create and send \acron response
& & \\
$response := app || \chal' || \er_{min} || \er_{max} || \auth$ & \sendmessageright{top={\acron response}} &  7) In \tcb-Wait: Authenticate the response, producing a one-bit output:

\begin{equation*}
	out := \authen(\attkey, \acron \text{ response})
\end{equation*} \\

& &
Based on $out$ and $app$, it decides the next transition:
\begin{compactitem}
	\item If $out=0$: Re-enter \tcb-Wait. {\it Jump to Step 2}.
	\item Else if $app=0$: Save ($\chal'$, $\er_{min}$ and $\er_{max}$ to $METADATA$) and enter \textit{\tcb-Heal}. {\it Jump to Step 8}.
	\item Else: Save ($\chal'$, $\er_{min}$ and $\er_{max}$) to $METADATA$, exit \tcb and resume execution of $\er$. {\it Jump to Step 9}. 
\end{compactitem}~\\

& & 8) In \tcb-Heal: Execute remediation software (e.g., reboot, reset, software update), then re-start \tcb-Att. {\it Jump to Step 1}. \\

& & \\

& & 9) Resume Application Execution:
    \begin{compactitem}
    \item Whenever executing $\er$: append control-flow transfers to \log.
    \item Whenever a \trigger occurs, \acron causes execution to enter \tcb-Att. {\it Jump to Step 1}.
    \end{compactitem}\\

\hline
\end{tabular}
\vspace{-1em}
\caption{\acron protocol.}
\label{fig:prot}
\vspace{-1.5em}
\end{figure*}